\documentclass[epj]{svjour}
\usepackage{graphics}
\usepackage{color}

\def\Pe{{\sf Pe}}
\def\calO{{\cal O}}

\def\a{\alpha}

\def\d{\delta}

\def\r{{\rho}}

\def\z{{\zeta}}

\def\br{\vec{r}}
\def\bs{\vec{\sigma}}

\def\bv{\vec{v}}

\def\bpi{\vec{\pi}}

\def\rB{{\rm{B}}}

\def\rg{{\rm{g}}}

\def\Rm{{\rm{m}}}

\def\ro{{\rm{o}}} 
\def\rp{{\rm{p}}}

\def\rs{{\rm{s}}}

\def\rT{{\rm{T}}}

\def\sn{{\rm{sn}}}

\def\tt{{\tilde{t}}}
\def\tx{{\tilde{x}}}

\makeatletter
  \def\mathcomposite{%
     \@ifstar
        {\def\@mathcomposite@option{%
            \baselineskip\z@skip\lineskiplimit-\maxdimen}%
         \@mathcomposite}%
        {\let\@mathcomposite@option\offinterlineskip
         \@mathcomposite}}
  \def\@mathcomposite{%
     \@ifnextchar[\@@mathcomposite{\@@mathcomposite[0]}}
  \def\@@mathcomposite[#1]#2#3#4{%
     #2{\mathchoice
        {\@mathcomposite@{#1}{#3}{#4}\displaystyle{1}}%
        {\@mathcomposite@{#1}{#3}{#4}\textstyle{1}}%
        {\@mathcomposite@{#1}{#3}{#4}%
         \scriptstyle\defaultscriptratio}%
        {\@mathcomposite@{#1}{#3}{#4}%
         \scriptscriptstyle\defaultscriptscriptratio}}}
  \def\@mathcomposite@#1#2#3#4#5{%
     \vcenter{\m@th\@mathcomposite@option
        \dimen@\f@size\p@\dimen@#1\dimen@\dimen@#5\dimen@
        \divide\dimen@ 18
        \edef\@mathcomposite@skipamount{\the\dimen@}%
        \ialign{\hfil$#4##$\hfil\cr
           #2\crcr
           \noalign{\vskip\@mathcomposite@skipamount}%
           #3\crcr}}}
\makeatother

\def\lsim{%
\hspace{0.3em}\raisebox{0.4ex}{$<$}\hspace{-0.75em}%
\raisebox{-.7ex}{$\sim$}\hspace{0.3em}%
}

\def\gsim{%
\hspace{0.3em}\raisebox{0.4ex}{$>$}\hspace{-0.75em}%
\raisebox{-.7ex}{$\sim$}\hspace{0.3em}%
}%
                         
\begin{document}
\title{How Flow Changes Polymer Depletion in a Slit}
\author{
Takashi Taniguchi\inst{1} 
\and 
Yuichiro Arai\inst{1} 
\and
Remco Tuinier\inst{2,3} 
\and 
Tai-Hsi Fan\inst{4}
}
\offprints{Takashi Taniguchi}
\mail{taniguch@cheme.kyoto-u.ac.jp}
\institute{
Graduate School of Engineering, Kyoto University 
Katsura Campus, Nishikyo-ku, Kyoto 615-8510, JAPAN
\and 
DSM ChemTech, ACES, P.O. Box 18, 6160 MD Geleen, The Netherlands
\and
Van~'t Hoff Laboratory for Physical and Colloid Chemistry, 
Debye Institute, Utrecht University, The Netherlands
\and
Department of Mechanical Engineering, 
University of Connecticut, CT 06269, USA
}
\date{Received: date / Revised version: date}
\abstract{
A theoretical model is developed 
for predicting dynamic polymer depletion 
under the influence of fluid flow. 
The results are established by combining the two-fluid model 
and the self-consistent field theory. 
We consider a uniform fluid flow across a slit containing a solution 
with polymer chains. 
The two parallel and infinitely long 
walls are permeable to solvent only 
and the polymers do not adsorb to these walls.
For a weak flow and a narrow slit,  
an analytic expression is derived 
to describe the steady state polymer concentration profiles
in a $\Theta$-solvent.
In both $\Theta$- and good-solvents, 
we compute the time evolution of the concentration profiles 
for various flow rates characterized by the Peclet number. 
The model reveals the interplay of depletion, 
solvent condition, slit width, and 
the relative strength of the fluid flow.
}
                           \maketitle

\section{Introduction}\label{sec:Introduction} 

In a polymer solution near an interface the polymer segments 
are either attracted or repelled by that interface \cite{Fleer1993}. 
In the latter case there exists a depletion zone near the interface. 
In this zone the polymer segment concentration is smaller 
than the bulk value because of 
the less possible number of configurations 
of the polymer chains.
The non-adsorbing wall 
forbids a certain amount of paths of the polymer chain. 
For ideal polymer chains near a hard wall the depletion thickness 
is close to the polymer's radius of gyration \cite{Eisenriegler1983}.  
This result holds generally 
for a dilute polymer solution \cite{Hanke1999,Fleer2008}, 
whereas in a semi-dilute polymer solution 
the depletion thickness is determined 
by the correlation length \cite{deGennes_scaling_concept1979}. 
These results for dilute and semi-dilute concentrations 
can also be combined \cite{Fleer2003,Fleer2007}. 
%
%
%
%
The previous investigations of polymer depletion 
at an interface primarily focus on the equilibrium case.
The equilibrium depletion thickness suffices to describe the attraction 
between colloidal particles 
when the depletion layers overlap \cite{Asakura1954,Vrij1976},
which can be measured for instance by optical tweezers \cite{Verma2000}.
The depletion force may yield 
phase transitions \cite{Lekkerkerker1992,Ilett1995,Meijer1994}
for which the binodals can be predicted 
for well-defined colloid-polymer mixtures \cite{Fleer2008,Tuinier2008}.
It is of fundamental and practical interest
to understand 
the change of the depletion layer under a fluid flow effect.  
Simple shear flow of a polymer solution next to a single wall 
leads to a slip-like behavior
even if the depletion layer is assumed 
to be unaffected by shear \cite{Tuinier2005}. 
It has been shown, however, 
that the flow does change the depletion thickness at a wall \cite{Duering1990}.
In colloidal systems,
the models developed for single particle motion
and pairwise particle interactions neglect the slight distortion 
of the depletion zone \cite{Tuinier2006,Fan2007,Fan2010}. 
This is applicable for describing long-time Brownian diffusion 
with weak convective effect. 
A convective depletion model was first established by Odijk 
for describing a thin depletion boundary layer in front 
of a fast moving sphere \cite{Odijk2004}. 
However, a complete picture of the fully coupled convective depletion effect
under various solution conditions 
is not yet available. 
%
%

In this paper we propose a theoretical framework 
to investigate dynamic depletion effects 
by combining two models : 
     (i)  the two-fluid model \cite{Doi1992}, 
          with the chemical potential obtained under 
          the ground state approximation (GSA) of 
          the self-consistent field theory
	  (SCFT) for polymeric systems
	  \cite{Fleer1993,deGennes_scaling_concept1979} 
and  (ii) the two-fluid model along with 
          the dynamical version (DSCFT) \cite{Hall2007}
	  of SCFT. 
To demonstrate how the combined formalism works well,
here we study 
the influence of flow on the polymer segment density profile 
in a narrow or wide slit. 
This is an interesting problem since for a narrow slit 
an analytic expression is only available 
for describing the equilibrium segment density profile 
\cite{Fleer1993,Fleer2003}.
It is unclear that to what extent 
a flow field modifies the depletion layers. 
Flow through pores is relevant for instance
in size-exclusion chromatography,
which is widely used to analyze polymers\cite{Pore}.

The content of this paper is as follows.
In the next section,
we start from a general formulation of the two-fluid model 
for polymer transport
to resolve the interplay of convective and diffusive effects.  
Then,
in Sec. \ref{subsec:Chemical Potential of Polymers}
we describe 
the ground state approximation of SCFT 
and the dynamical version of SCFT 
to express the chemical potential which 
characterizes the segment density profiles
under various flow conditions.
%
%
%
%
In Sec. \ref{subsec:One-dimentional Formulation},
we describe a set of equations to investigate 
the dynamics of polymer segment depletion in a slit.
In Sec. \ref{sec:Results and Discussion},
the results for both $\Theta$- and good-solvent conditions
are provided in detail.
Finally, we give a conclusion in Sec. \ref{sec:Conclusions}.
%
%

\section{Theory}\label{sec:Theory}

Here we describe theoretical models 
to investigate dynamic depletion effects under a flow.
The models used here are combinations of the two fluid model 
with a model to evaluate the chemical potential for polymeric systems,
{\em i.e.},
  (i) the two-fluid models with a chemical potential 
  evaluated under GSA of SCFT 
  and 
  (ii) two fluid model with DSCFT. 
%
%
The validity for each model has been reported in literature. 
The two-fluid model of polymeric materials was developed 
by Doi and Onuki \cite{Doi1992}. 
The model has succeeded in explaining the viscoelastic
behaviors \cite{R1,R2} and shear induced phase separation
in polymer solutions and polymer blends \cite{R3}. 
The validity of the model has been confirmed 
by simulations \cite{R2,R3} and experiments \cite{R1,R4},
and thus is suitable to describe non-equilibrium transport phenomena 
in polymeric systems. 
The SCFT is frequently used for 
predicting the equilibrium properties 
of inhomogeneous polymeric systems. 
It gives a precise evaluation of chemical potential for 
each constituent 
in the system \cite{Fleer1993,deGennes_scaling_concept1979}. 
%
%
%
%
%
%
In the two-fluid model,
the determination of the local chemical potential of polymer segment is
critical especially for the reduction of the possible chain conformations 
near the wall (the depletion effect). 
Therefore, SCFT theory is more suitable than Flory-Huggins
theory which was developed for evaluating bulk properties.
The computational costs of SCFT are 
higher than using the Flory-Huggins theory 
because the statistical weight of each polymer conformation 
must be evaluated by solving a diffusion-like differential equation 
unless the ground state approximation is used, 
which is applicable when the gyration radius of 
the constituent polymers is sufficiently large compared 
to the length scale of the confined space. 
At equilibrium, 
the GSA for the depletion layer near a flat wall agrees well with 
the numerical and Monte Carlo simulation results \cite{RemcoBook2011,R5}.
When the system is out of equilibrium under a constant flow condition, 
a dynamical version of SCFT (DSCFT) is 
needed to study the polymer depletion effect. 
The validity and efficiency of DSCFT for various cases
has been reported in the literature \cite{Hall2007}.

In the following subsections, we briefly explain the essence of 
the two-fluid model, dynamical version of the self-consistent field theory and
the ground state approximation applied to the SCFT.

\subsection{Two-Fluid Model}
\label{subsec:two-fluid model}

We consider inertialess fluid motion
and polymers in dilute to semi-dilute polymer solutions.
The polymer solution consists of solvent and 
homopolymer with length $aN$, 
where $a$ and $N$ are the monomer size and the polymerization index,
respectively.
The transient-evolution of polymer segment volume fraction $\phi_\rp$
is given by the continuity equation :
\begin{eqnarray}
{\partial \phi_\rp(\br,t) \over \partial t} 
= - \nabla \cdot (\phi_\rp \bv_\rp), 
\label{eqn:eq_of_continuity}
\end{eqnarray}
where 
$t$ is time, 
$\bv_\rp$ is the velocity of 
the polymer segments in the fluid,  
$\phi_\rp(\br,t) = a^3 \r_\rp(\br,t)$ 
with $a$ being the segment length  
and $\rho_\rp$ the local number density of polymer segments. 
The local volume fraction of solvent is $\phi_\rs$.
Because $\phi_\rp + \phi_\rs = 1$, 
the total velocity $\bv(\br,t)=\phi_\rp \bv_\rp  + \phi_\rs \bv_\rs$
($\bv_\rs$ is the solvent velocity)  
satisfies the incompressibility condition $\nabla \cdot \bv = 0$.
The momentum equation of the two-fluid model can be derived
from the Rayleighian given by Doi and Onuki \cite{Doi1992}: 
\begin{eqnarray}
{\cal R}= \int 
\biggr [ && 
{\z(\br,t) \over 2 } | \bv_\rp - \bv_\rs |^2 
- \mu_\rp \nabla \cdot (\phi_\rp \bv_\rp)
- \mu_\rs \nabla \cdot (\phi_\rs \bv_\rs)
\nonumber \\
&& 
- p \nabla \cdot \bv
+ \bs_\rp : \nabla \bv_\rp 
+ \bs_\rs : \nabla \bv_\rs 
\biggr ]d\br,
\label{eqn:Rayleighian}
\end{eqnarray}
where 
$\mu_\rp$, 
$\mu_\rs$, 
$\bs_\rp$,
$\bs_\rs$,
and 
$p$ 
are the local chemical potential of polymer and solvent,
    the deviatoric stress of polymer and solvent,  
and the hydrodynamic pressure,
respectively. 
The friction coefficient $\z$ 
between the two fluids can be 
approximated as Stokes friction coefficient for the polymer blob
per volume \cite{deGennes_scaling_concept1979}, 
$\zeta = 6 \pi \eta_\rs \xi/\xi^3 = 6\pi\eta_\rs / \xi^2$,
where $\xi(\phi_\rp)$ is the blob size 
and $\eta_\rs$ is the solvent viscosity. 
The blob size is related to $\phi_\rp$ as 
$\xi \simeq a \phi_\rp^{-m}$ \cite{deGennes_scaling_concept1979},
where $m=1$ and $3/4$ for $\Theta$- and good solvents, respectively.
By minimizing $\cal R$ with respect to $\bv_\rp$ and $\bv_\rs$, 
the following momentum equations can be derived : 
%
\begin{eqnarray}
&& 
   \z (\bv_\rp-\bv_\rs) 
  + \phi_\rp \nabla p 
  + \phi_\rp \nabla \mu_\rp
  - \nabla \cdot \bs_\rp
= 0, 
\label{eqn:vp}
\\ 
&& 
    \z (\bv_\rs-\bv_\rp) 
  + \phi_\rs \nabla p 
  + \phi_\rs \nabla \mu_\rs
  - \nabla \cdot \bs_\rs
= 0. 
\label{eqn:vs}
\end{eqnarray}
%
By combining Eqs.(\ref{eqn:vp}) and (\ref{eqn:vs}) we obtain 
%
%
\begin{eqnarray}
  \nabla \cdot \bs
 -\nabla P 
 - \nabla \cdot \vec{\pi}
= 0, &&
\label{eqn:v}
\end{eqnarray}
%
%
where the total stress $\bs=\bs_\rp+\bs_\rs$, 
$P$ is the modified pressure defined by $P=p+\mu_\rs$, 
and $\vec{\pi}$ is the osmotic pressure tensor defined as 
\begin{equation}
\nabla \cdot \vec{\pi} = \phi_\rp(\br,t) \nabla \mu(\br,t),   
\label{eqn:div_pi}
\end{equation}
where $\mu=\mu_\rp-\mu_\rs$ is 
the difference between the chemical potentials.
Substituting Eq.(\ref{eqn:v}) into Eq.(\ref{eqn:vs}) 
and eliminating $P$, 
the following expression for the polymer velocity is obtained:
%
%
\begin{eqnarray}
  \bv_\rp 
\simeq  && 
\bv 
 - {\phi_\rs \over \z } \biggr [
   \phi_\rs \bigr ( 
                    \nabla \cdot \vec{\pi}
                 -  \nabla \cdot \bs_\rp
           \bigr )
                 + \phi_\rp \nabla \cdot \bs_\rs
            \biggr ].
\label{eqn:vp3}
\end{eqnarray}
%
%
In this paper we consider $\phi_\rp\ll 1$  
for dilute and semi-dilute solutions, 
and $\phi_\rp \nabla \cdot \bs_\rs $ of the above equation is negligible.
The polymer velocity relative to the total velocity is 
driven by the osmotic pressure and the deviatoric stress 
of the polymer fluid.
Substituting Eq.(\ref{eqn:vp3}) into Eq.(\ref{eqn:eq_of_continuity})
yields the polymer transport equation : 
%
%
\begin{eqnarray}
	{\partial \phi_\rp \over \partial t} 
+ \nabla \cdot (\phi_\rp \bv) 
\simeq \nabla \cdot 
\biggr [ { \phi_\rp \phi_\rs^2 \over \z(\phi_\rp) }
  \biggr (
  \phi_\rp \nabla \mu - \nabla \cdot \bs_\rp
  \biggr )
\biggr ].
\label{eqn:eq_of_continuity2}
\end{eqnarray}
%
%
This formulation is consistent with Odijk's work \cite{Odijk2004} 
except for the $\phi_\rp$-dependent friction coefficient
and the additional $\nabla \cdot \bs_\rp$ term. 
For a solid object with an arbitrary shape 
$\Omega$ with a surface $\partial \Omega$, 
the osmosis-induced force and torque are 
$\vec{f} = \int_{\partial \Omega} 
                 (- \vec{\pi} \cdot \vec{n})
                  dS $
and 
$\vec{T} = \int_{\partial \Omega} 
                  \vec{R}\times 
                  (-\vec{\pi} \cdot \vec{n}) dS $,
respectively, 
where 
$\vec{n}$ is the surface normal and  
$\vec{R}$ is the vector  
from the center of gravity 
to the surface of the object. 
\par
When assume that 
the polymer stress in the dilute and semi-dilute 
($\bar\phi_\rp \lsim$ overlap volume 
fraction $\phi_\rp^\ast \simeq N^{1-3\nu}$) 
polymer solution can be expressed as 
$ \eta' \phi_\rp \bigr (\nabla \bv +{}(\nabla \bv)^\rT \bigr )$,  
where $\eta'$ is a constant.
Hence the total velocity follows as $\bv \sim {\cal O}( {\bar\phi}_\rp^2 )$, 
as seen from Eq.(\ref{eqn:v}) and $\bpi \sim {\cal O}(\bar\phi_\rp^2)$, 
and thus $\nabla\cdot\bs_\rp \sim {\cal O}(\bar\phi_\rp^3)$.
Also because $\mu \sim {\cal O}({\bar\phi}_\rp)$, 
so $ \phi_\rp\nabla\mu_\rp \sim \calO(\bar\phi_\rp^2)$, 
this implies 
$  \calO( \nabla\cdot\bs_\rp ) \ll \calO( \phi_\rp\nabla\mu  )$ and 
the polymer force $\nabla \cdot \bs_\rp$ in Eq.(\ref{eqn:eq_of_continuity2}) 
is indeed negligible \cite{Odijk2004}.
%

\subsection{Chemical Potential of Polymers}
\label{subsec:Chemical Potential of Polymers}

\subsubsection{Self-Consistent Field Theory}

Using SCFT, the Helmholtz free energy is written as 
\begin{eqnarray}
{F \over k_\rB T}= 
{1 \over a^3} \int 
&&
\biggr [ 
\chi \phi_\rp \phi_\rs  - w_\rp\phi_\rp - w_\rs\phi_\rs 
\biggr ] d\br
\nonumber\\
&& 
- {\bar \phi_\rp \over N } \ln ( { N Q_\rp \over {V \bar \phi_\rp}  } ) 
- {\bar \phi_\rs}          \ln ( {   Q_\rs \over {V \bar \phi_\rs} } ), 
\label{eq:Helmholtz2}
\end{eqnarray}
where $N$ is the number of segment in a single chain, 
$V$ is the volume of the system, and 
$w_\a$ and $\bar \phi_\a$ are 
the local dimensionless interaction field
(scaled by $k_\rB T$) 
and the average volume fraction of polymer segments in the bulk
for the $\a$-component ($\a$ represents polymer or solvent), 
respectively. 
The partition functions  $Q_\rp$ and $Q_\rs$ 
for a single chain and a single solvent, respectively,  
are defined by
\begin{eqnarray}
Q_\rp = && \int_V d\br \int_0^1 q(\br,s)q(\br,1-s) ds, 
\label{eq:Q_p}
\end{eqnarray}
and 
\begin{eqnarray}
Q_\rs = && \int_V d\br \exp(-w_\rs(\br)).
\label{eq:Q_s}
\end{eqnarray}
In Eq.(\ref{eq:Q_p}), 
$q$ is the statistical weight of the polymer chain
and can be calculated by 
\begin{eqnarray}
{\partial q \over \partial s } = 
 {a^2 \over 6 } N \nabla^2q - N w_\rp(\br)q  
\quad  {\rm for}\quad  0 \le s \le 1
\label{eqn:q}
\end{eqnarray}
with the initial condition $q(\br,0) = 1$
and the zero boundary condition at the solid surface 
$q(\br\in \partial \Omega) = 0$ 
for an arbitrary contour coordinate $s$.

In order to evaluate $\mu_\rp$ and $\mu_\rs$ for a given $\phi_\rp(\br,t)$, 
the following iterative procedure can be used \cite{Hall2007}:

\noindent
i)  Find $w_\rp(\br)$ and $q(\br,s)$ that 
    satisfy Eq.(\ref{eqn:q}) and  
    \begin{eqnarray}
    { 1 \over Q_\rp } \int_0^1 q(\br,s)q(\br,1-s) ds = 
    {1 \over V \bar\phi_\rp} \phi_\rp(\br,t) 
    \label{eq:wp}
    \end{eqnarray}
along with the initial and boundary conditions.

\noindent
ii) Evaluate $w_\rs(\br)$ 
    that satisfies 
    \begin{equation}
    { 1 \over Q_\rs } \exp(-w_\rs(\br)) 
    = { 1 \over V\bar\phi_\rs } \phi_\rs(\br,t).
    \label{eq:ws} 
    \end{equation}

\noindent
iii) Compute the chemical potential difference $\mu=\mu_\rp-\mu_\rs$ 
     from the functional derivatives:
      \begin{eqnarray}
      \mu_\rp(\br) = && { \d F \over \d \phi_\rp }
                  = - w_\rp(\br) + \chi \phi_\rs(\br), 
      \label{eq:mu_p} \\
      \mu_\rs(\br) = && { \d F \over \d \phi_\rs }
                  = - w_\rs(\br) + \chi \phi_\rp(\br).
      \label{eq:mu_s}
      \end{eqnarray}

\noindent
iv) Finally, the time evolution of $\phi_\rp(\br,t)$ 
    can be calculated by Eq.(\ref{eqn:eq_of_continuity2})
    with $\bs_\rp=\vec{0}$.

\subsubsection{Ground State Approximation}

In a confined space (Fig.\ref{fig:Fig_01})
we can apply the ground state approximation (GSA) 
to the self-consistent field theory (SCFT).  
The Helmholtz free energy of the mixture 
can be expressed as \cite{deGennes_scaling_concept1979}
\begin{eqnarray}
F = \epsilon_\ro
\int \biggr [   \ell^2 (\nabla \sqrt{ \phi_\rp} )^2 
               + f_\ro(\phi_\rp)
                 \biggr ] d\br,  
\label{eqn:FreeEnergy} 
\end{eqnarray}
where ${\epsilon_\ro = k_\rB T / a^3 }$, 
$\ell^2 = {a^2 / 6}$ in GSA or 
$\ell^2 = {a^2 / 9}$ in the random phase approximation (RPA), 
$f_\ro(\phi_\rp) = 
                  (1-\phi_\rp) \ln (1-\phi_\rp) 
                + \chi \phi_\rp (1-\phi_\rp),
$
and $\chi$ is the Flory-Huggins parameter. 
The reason why a translational entropy term 
$\phi_\rp/N \ln \phi_\rp$ 
does not appear in $f_\ro$ has been explained 
by Fredrickson \cite{BookOfFredrickson}.
The chemical potential difference $\mu$ is determined by 
the functional derivative of $F$ with respect to $\phi_\rp$:   
\begin{eqnarray}
\mu(\br) 
= \epsilon_\ro \biggr [
             - { \ell^2 \over \sqrt{\phi_\rp} }
             \nabla^2 \sqrt{\phi_\rp}
           + (1-2\chi)\phi_\rp 
            \biggr ]. 
\qquad
\label{eqn:ChemicalPotential}
\end{eqnarray}
Since $\phi_\rp$ is much smaller than unity, 
only the zeroth- and first-order terms involving $\phi_\rp$
in the chemical potential are preserved.
The corresponding osmotic stress tensor $\pi_{ij}$,  
scaled by $\epsilon_\ro$, 
can be expressed as 
\begin{eqnarray}
&& 
\hspace{-15mm}
\pi_{ij} = \biggr \{ \Bigr [ \pi_\ro
- \ell^2 \nabla \cdot \bigr ( \sqrt{\phi_\rp} \nabla \sqrt{\phi_\rp}  
                          \bigr )
                                     \Bigr ] \delta_{ij} 
\nonumber\\
&& \qquad \qquad \qquad + 2\ell^2 
{\partial \sqrt{\phi_\rp} \over \partial x_i}
{\partial \sqrt{\phi_\rp} \over \partial x_j}
\biggr \}, 
\label{eqn:osmotic_stress}
\end{eqnarray}
where $\pi_\ro(\phi_\rp)=\phi_\rp f'_\ro(\phi_\rp) - f_\ro$
is the osmotic pressure in the bulk.
Note that $\nabla \cdot \vec{\pi}$ 
satisfies Eq.(\ref{eqn:div_pi}).
%
%

\subsection{One-dimensional Formulation}
\label{subsec:One-dimentional Formulation}

\begin{figure}
\begin{center}
\resizebox{1.00\columnwidth}{!}{\includegraphics{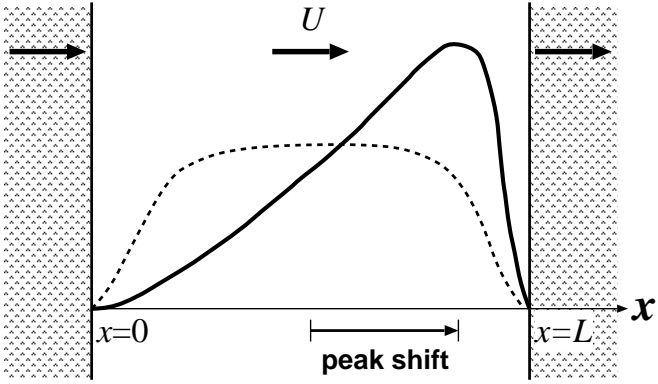}}
\end{center}
\caption{
A schematic illustration of the convective depletion effect 
on the polymer segment profile (solid line) 
in a slit between two parallel and solvent permeable walls. 
The dashed line indicates the equilibrium distribution.
The polymer concentration vanishes at the walls. 
}
\label{fig:Fig_01}
\end{figure}

%
%
We focus on non-equilibrium 
polymer segment 
concentration profiles 
under a uniform flow $U$ 
passing through two parallel and solvent-permeable walls 
separated by a distance $L$ (Fig.\ref{fig:Fig_01}). 
%
%
%
%
There are three length scales
involved in this problem,  
{\em i.e.,}
the width $L$, 
the monomer size $a$, and 
gyration radius $R_\rg$ of the ideal polymer chain,  
defined as $R_\rg = a(N/6)^{1/2}$.
We select $L$ as the length scale, 
and $\tau \equiv L^2/D$ as the time scale, 
where $D$ is a diffusion coefficient defined 
by $D = {\bar \phi_\rp}^{1-2m} D_{\rm self}$,  
where $D_{\rm self}={  k_\rB T / 6\pi \eta_\rs a  }$ is
the self-diffusion coefficient of a single polymer segment.
It should be noted that 
the factor $\bar \phi_\rp^{1-2m}$ 
comes from the fact that a polymer segment is not affected 
directly by hydrodynamic flow, 
but a blob with a size $\xi$ as a whole is affected 
by hydrodynamic flow, 
in a way that the friction constant was introduced as
$\zeta(\phi_\rp) = 6\pi\eta_\rs \xi/$[blob volume].
The volume fraction of polymer 
is scaled by the averaged volume fraction $\bar \phi_\rp$, 
and the chemical potential is scaled by 
$k_\rB T / a^3$. 
Hereafter we use dimensionless expressions.
The scaled chemical potential difference is expressed as
\begin{equation}
\mu
          = - {\ell^2 \over L^2 }{1 \over \varphi }
           {\partial^2 \varphi \over \partial x^2 }
           + {v} \varphi^2, 
\label{eqn:ChemicalPotential_2}
\end{equation}
where 
$\ell = (a^2/6)^{1/2}$, 
$v=(1-2\chi)\bar\phi_\rp$
and $\varphi = (\phi_\rp)^{1/2}$. 
Note that the ground-state approximation, 
Eq.(\ref{eqn:ChemicalPotential_2}) 
is valid only for a narrow slit $L/R_\rg<\pi$ 
\cite{deGennes_scaling_concept1979}.
The polymer transport equation (Eq.(\ref{eqn:eq_of_continuity2}))
thus can be expressed as 
\begin{eqnarray}
      {\partial \varphi^2 \over \partial t} 
    + {\sf Pe} {\partial \varphi^2 \over \partial x }
    = 
      {\partial \over \partial x} 
      \biggr [ 
      \varphi^{2(2-2m)} {\partial \mu \over \partial x}
      \biggr ], 
\label{eqn:phi_2}
\end{eqnarray}
where 
${\sf Pe}=U L / D$ 
is the Peclet number and 
the width of the slit $L$ is the characteristic length of the system. 
The control parameters are ${\sf Pe}$, $v$, and $L/\ell$. 
The translational entropy term containing the polymerization index $N$
only appears in a wide slit case ($L \gg R_\rg$), but not 
in the transport equation, 
eq.(\ref{eqn:phi_2}) with eq.(\ref{eqn:ChemicalPotential_2}), 
for the narrow slit case under GSA. 
Since the polymer solution we consider consists of homopolymer and solvent,  
the monomer size $a$ is not a parameter but a fixed constant.
Hence, using $R_\rg \equiv \ell N^{1/2}$, 
the intrinsic control parameters in the present system 
are found to be $\Pe, v,$ and $L/R_\rg.$
It should be noted that the average volume fraction 
$\bar \phi_\rp$ is included implicitly 
in $v$ as $v =(1-2\chi)\bar \phi_\rp$.
In the following numerical calculations, 
we applied various values, 
the Flory parameter $\chi$ to realize variations
in the excluded volume parameter $v$. 
The value of $\bar\phi_\rp$ is selected to be 
close to or less than the overlap volume fraction $\phi_\rp^\ast$ 
because the viscoelastic effect 
($\bs_\rp$ in eq.(\ref{eqn:eq_of_continuity2})) is neglected here.
When the average volume fraction is much smaller than $\phi_\rp^\ast$, 
the system gives almost the same behavior as $\Theta$-solvent 
due to the effectively small excluded volume effect ($v \simeq 0$). 
From the above reason, we consider $\bar \phi_\rp \simeq \phi_\rp^\ast$.

By applying a constant flow with uniform velocity $U$ 
at $t=0$
to a quiescent polymer solution in a slit, 
polymer segments start to accumulate
at a region near the wall on the downstream side
and then the system reaches a steady state 
with a region of accumulated polymer segments,
characterized by a peak at which the polymer segment
concentration passes through a maximum.
The corresponding convective or accumulation time scale can be estimated by
$t^\ast=L/U$. 
Hereinafter we refer to it as "accumulation time''. 
So, the dimensionless accumulation time 
is ${\tilde t}^\ast (\equiv t^\ast/\tau)$ 
$\sim 1/{\sf Pe}$.
Note that in the estimation of $t^\ast$
the distance of the peak shift is less than $L$  
and the effective advection velocity 
is probably smaller than $U$
due to the thermodynamic flux 
which tends to reduce concentration gradients.

Because the walls are not permeable to polymers,
the corresponding boundary conditions 
are $\varphi(0)$=0 and $\varphi(1)$=0, 
and the zero flux boundary conditions 
$d\mu/dx{|}_{x=0}$=0 
and ${d\mu / dx}{|}_{x=1}$=0
are applied.
Equation (\ref{eqn:phi_2}) leads to the steady state solution:
\begin{eqnarray}
\mu(x) 
- 
\mu(0) 
= {\sf Pe}~  \int_0^x \varphi^{2(2m-1)}(x') dx'. 
\label{eqn:phi_steady_state} 
\end{eqnarray}

%
%
Under a weak flow condition, ${\sf Pe}\ll 1$,  
the first-order approximation of the steady state solution
for $\varphi$ and $\mu$ can be written as 
\begin{eqnarray}
&& \varphi(x) = \sqrt{\phi_\rp(x)} \simeq \varphi_\ro(x) 
             + {\sf Pe}~\varphi_1(x), 
\label{eqn:varphi}
\end{eqnarray}
and 
\begin{eqnarray}
&& \mu(x) \simeq \mu_\ro(x) + {\sf Pe}~\mu_1(x). 
\label{eqn:mu_upto_1st_order_in_Pe}
\end{eqnarray}
By substituting Eqs.(\ref{eqn:varphi}) 
and (\ref{eqn:mu_upto_1st_order_in_Pe})
into Eq.(\ref{eqn:phi_steady_state}), 
the corresponding leading and first-order equations become 
\begin{eqnarray}
{\ell^2 \over L^2}
  {d^2 \varphi_\ro \over dx^2}
- v \varphi_\ro^3
+ { \mu_\ro(0) \varphi_\ro } = 0, 
\label{eqn:zeroth_order_equation}
\end{eqnarray}
and
\begin{eqnarray}
&& \biggr [~   
{\ell^2 \over L^2}
{d^2  \over dx^2}
             + \mu_\ro(0) - 3 v \varphi_\ro^2 
  ~\biggr ] \varphi_1 
= 
\nonumber
\\ 
&& 
\quad
    - \varphi_\ro(x) \biggr [   \mu_1(0)  
                          + 
                             \int_0^x \varphi_\ro^{2(2m-1)}(x') dx'
                 \biggr ], 
\quad
\label{eqn:first_order_equation}
\end{eqnarray}
respectively.   
Both $\varphi_0$ and $\varphi_1$ vanish at the walls. 
Next we consider the results for 
$\Theta$- and good-solvent cases
under narrow and wide slit conditions.

\section{Results and Discussion}
\label{sec:Results and Discussion}

\subsection{$\Theta$-solvent Cases}

\subsubsection{Narrow Slit, $L < \pi R_\rg$}
\label{sec:Narrow_slit_theta_solvent}

The ground state approximation is valid 
when the distance between two walls is small, 
{\em i.e.,} $L < \pi R_\rg$.
Equation (\ref{eqn:ChemicalPotential_2}) 
with $v=0$ and Eq.(\ref{eqn:phi_2}) 
are used to investigate 
the time evolution of polymer segment concentration profile.
Under steady and weak flow conditions in $\Theta$-solvent ($m=1$, $\nu=0$), 
Eqs.(\ref{eqn:zeroth_order_equation}) 
and (\ref{eqn:first_order_equation})
reduce to 
\begin{eqnarray}
{\ell^2 \over L^2}
{d^2 \varphi_\ro \over dx^2}= - { \mu_\ro(0) \varphi_\ro },
\label{eqn:zeroth_order_equation:chi=0.5} 
\end{eqnarray}
and
\begin{eqnarray}
&& \biggr [ 
{\ell^2 \over L^2}
      {d^2  \over dx^2}
    + \mu_\ro(0) \biggr ] \varphi_1 
= 
\nonumber
\\ 
&& 
\qquad\quad
- \varphi_\ro(x) \biggr [ \mu_1(0) 
                          + 
                             \int_0^x \varphi_\ro^2(x') dx' 
                 \biggr ].
\qquad
\label{eqn:first_order_equation:chi=0.5}
\end{eqnarray}
\begin{figure}[t]
\begin{center}
\rotatebox{-90}{
\resizebox{0.65\columnwidth}{!}{\includegraphics{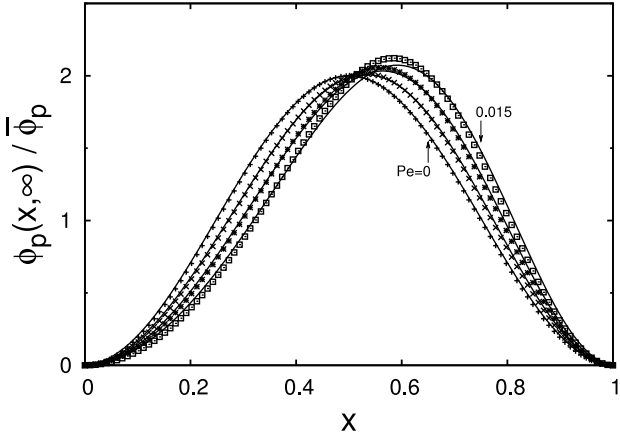}}}
\end{center}
\caption{
Steady state polymer segment volume fraction $\phi_\rp(x,t=\infty)$ 
of a polymer solution ($\bar \phi_\rp=0.1$, $N=1000$ and $\chi=0.5$) 
in a slit ($L/R_\rg=1$, where 
$R_\rg\simeq 33\ell$) at 
{\sf Pe}=0, 0.5, 1.0, and 1.5 $\times 10^{-2}$. 
The solid lines are 
from the numerical solutions of Eqs.(\ref{eqn:ChemicalPotential_2})
and (\ref{eqn:phi_2}).  
The symbols ($+$, $\times$, $\ast$, $\sqcup\!\!\!\!\sqcap$)
indicate the analytical approximations
for {\sf Pe}=0, 0.5, 1.0, and 1.5 $\times 10^{-2}$,
respectively. 
}
\label{fig:Fig_02_graph_1d}
\end{figure}
%
%
\begin{figure}
\begin{center}
\rotatebox{-90}{\resizebox{0.67\columnwidth}{!}{\includegraphics{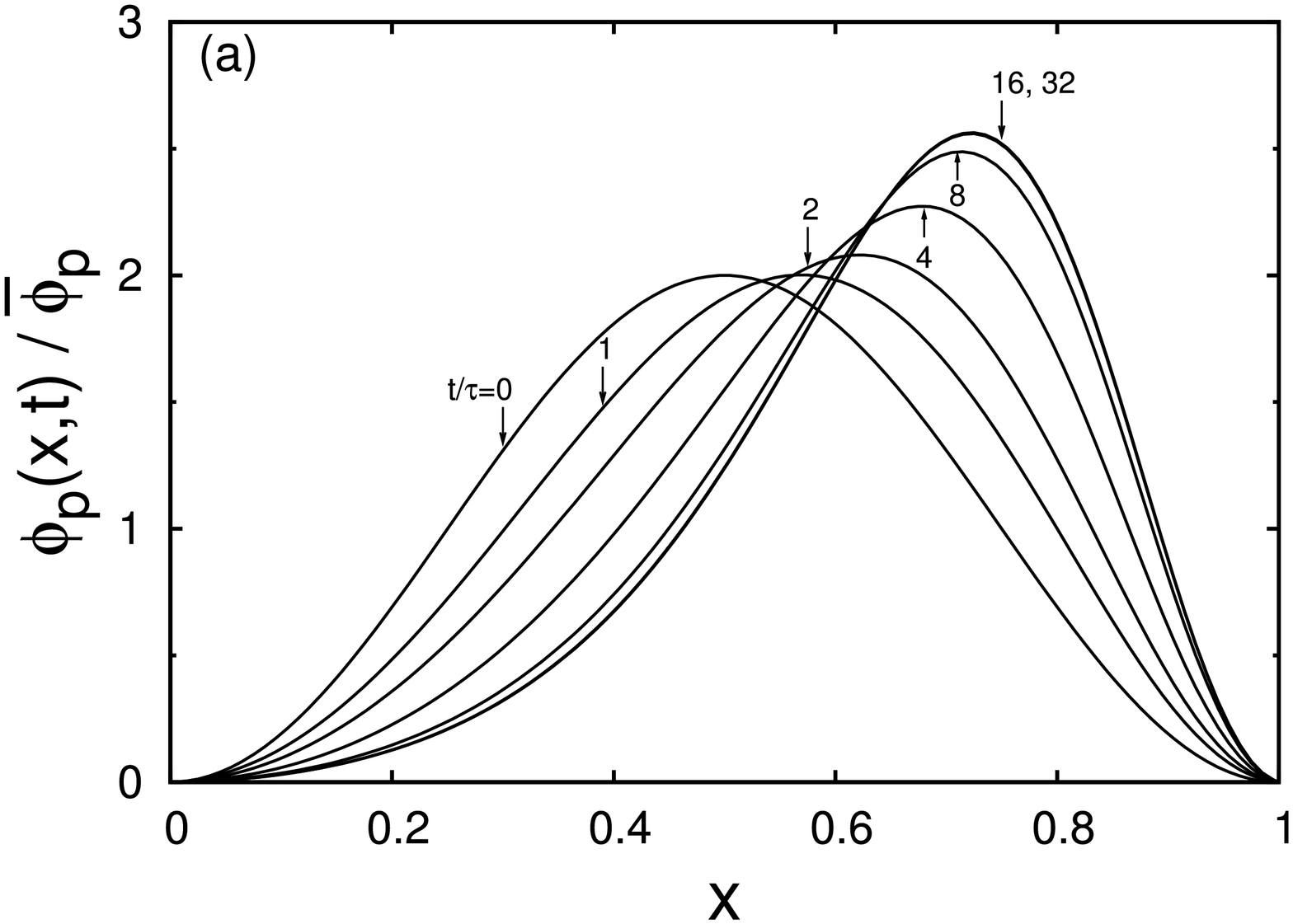}}}
\rotatebox{-90}{\resizebox{0.60\columnwidth}{!}{\includegraphics{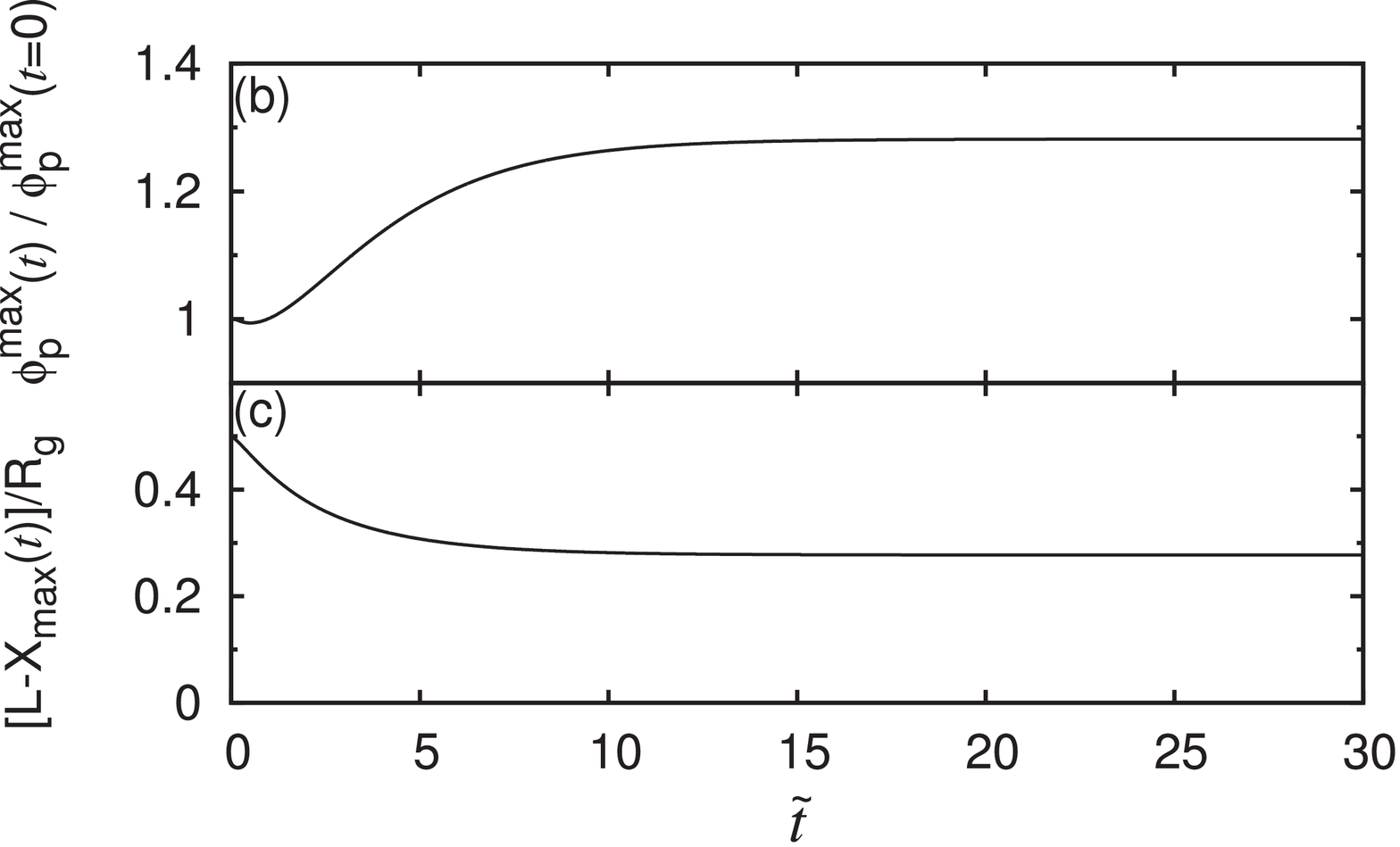}}}
\end{center}
\caption{
(a) Transient evolutions of the segment concentration profiles 
    $\phi_\rp(x,t)$,  
(b) the peak concentration and 
(c) the distance from the peak position 
    to the downstream-side wall.
Applied parameters are: $\bar \phi_\rp=0.1$, $N=1000$, $\chi=0.5$, 
$L/R_\rg=1$, and 
{\sf Pe}=0.05 at $\tilde t = t/\tau \ge 0$.
%
%
The profiles shown in (a) 
are at $\tilde t=$0 (no flow), 1, 2, 4, 8, 16 and 32.
At $\tt {{>}\atop{\sim}} 12$, 
the profile approaches a steady state as can be seen 
in panels (b) and (c). 
}
\label{fig:Fig_03_profile_time_evo_Pe=0.1_chi=0.5_phi0=0.01_Lx=1}
\end{figure}
%
%

\noindent
The zeroth-order solution corresponding 
to the equilibrium state is \cite{deGennes_scaling_concept1979} : 
\begin{eqnarray}
\varphi_\ro(x) 
= \sqrt{ \phi_\ro(x) }
= \sqrt{2}
\sin ( \pi x )
~~{\rm for}\quad 0\le x \le 1, 
\label{eqn:fundamental_sol}
\end{eqnarray}
which satisfies the normalization condition, $\int_0^1 \phi_\ro(x)dx$$=1$, 
and thus $\mu_\ro(0)=\pi^2 \ell^2 /L^2$.
The first-order solution reads
\begin{eqnarray}
\varphi_1(x)
&& 
   = A \sin ( \pi x  )
\nonumber\\
&&
  - B 
  \bigr \{ ~  
            ( 
               1 + 8 \pi^2 x 
                 - 8 \pi^2 x^2
            )
           \cos ( \pi x )
\nonumber\\
&&
\qquad~~
         - \cos ( 3 \pi x )
         + 4 \pi x \sin (   \pi x ) ~
  \bigr \},  \qquad
\label{eqn:varphi_solution}
\end{eqnarray}
%
\noindent
where the constant 
$B = L^2 / 16 \ell^2 \pi^3 \sqrt{2} $
and $\mu_1(0)$ in Eq.(\ref{eqn:first_order_equation:chi=0.5})
is found to be $-1/2$ by the boundary condition $\varphi_1(1)=0$. 
The constant $A$ can be determined by the normalization condition   
$\int_0^1 \phi(x)dx=1$, 
and we obtain 
\begin{eqnarray}
A && 
= 2\pi B + { B^2 \over \sqrt{2}} 
                 \Bigr ( 
                          77 - 2\pi^2 - {16 \over 15} \pi^4
                 \Bigr ){\sf Pe} 
         + {\cal O}({\sf Pe}^2). \quad
\label{eqn:s0_2}
\end{eqnarray}
Accordingly, the peak value and its location can be expressed as 
\begin{eqnarray}
{{\phi_\rp^{\rm max}({\sf Pe}) \over \phi_\rp^{\rm max}(0)}}
   && 
      = 1 + B^2 
          \Bigr ( 
                   77-2\pi^2+{14 \over 15}\pi^4
          \Bigr )         
          {\sf Pe^2}
        + \cdots
\label{eqn:peak_height}
\end{eqnarray}
and 
\begin{eqnarray}
   x_{\rm max} &&
    = 
            {  1 \over 2 } 
                    \Bigr [ 1
                        + 2 \sqrt{2} \pi B {\sf Pe} 
                        + {\cal O}({\sf Pe}^{3}) 
                    \Bigr ].
\label{eqn:peak_position}
\end{eqnarray}
From eq.(\ref{eqn:osmotic_stress}), 
the osmotic stresses acting on the walls at $x=0$ and $x=1$ are
\begin{eqnarray}
&&   \pi_{xx} = \epsilon_\ro 
                 \pi^2 {\bar \phi_\rp} { \ell^2 \over L^2} 
                 \biggr [ 
                          \sqrt{2}
                          \mp \big ( 6\pi B {\sf Pe} + C {\sf Pe}^2 \big )  
                 \biggr ]^2,
\end{eqnarray}
where "$-$" is for $x=0$ and "$+$" is for $x=1$, 
and $C=B^2(77-2\pi^2-16\pi^4/15)/\sqrt{2}$.
%
%
%
%
\begin{figure}
\begin{center}
\rotatebox{-90}{\resizebox{0.65\columnwidth}{!}{\includegraphics{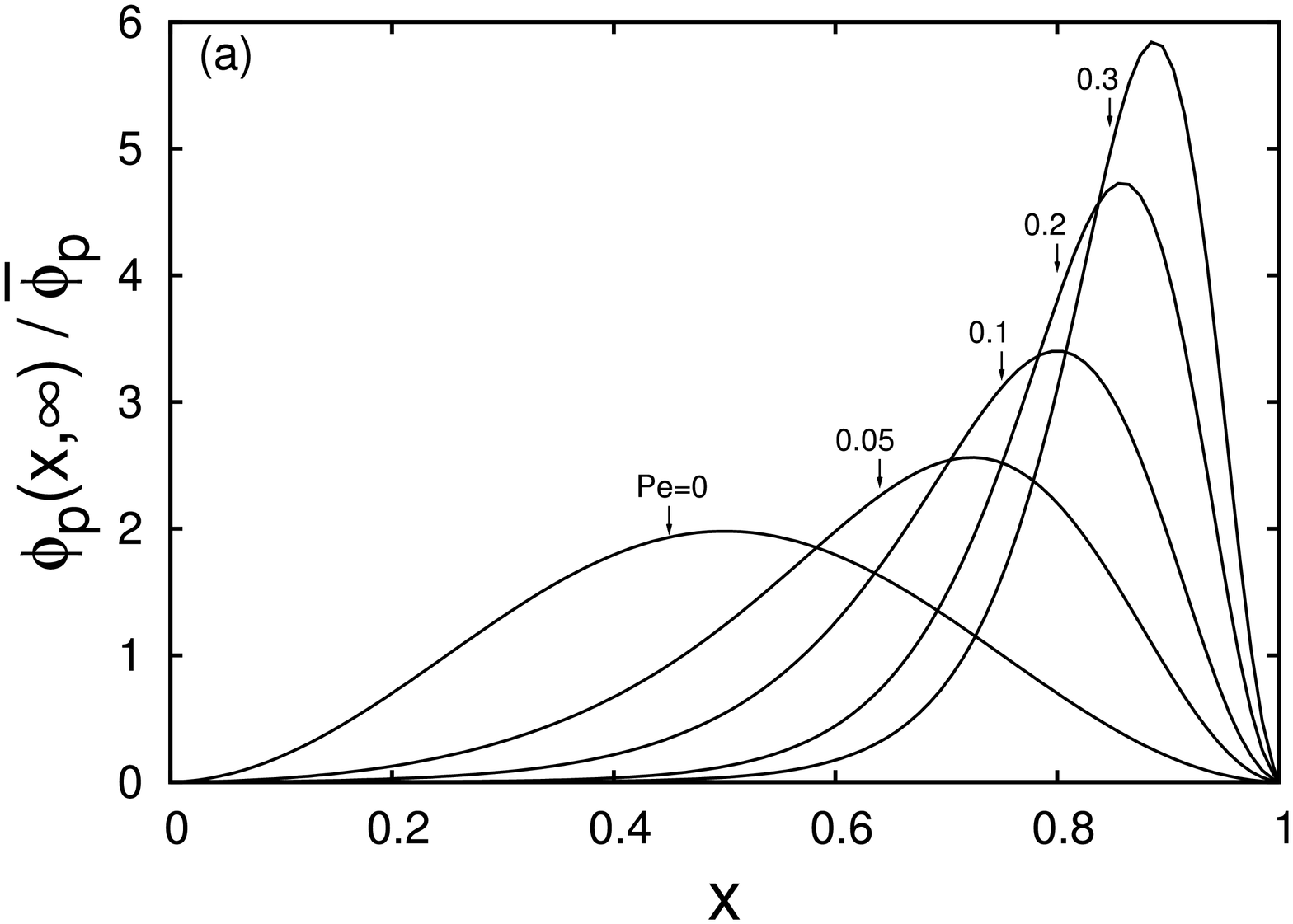}}}
\rotatebox{-90}{\resizebox{0.65\columnwidth}{!}{\includegraphics{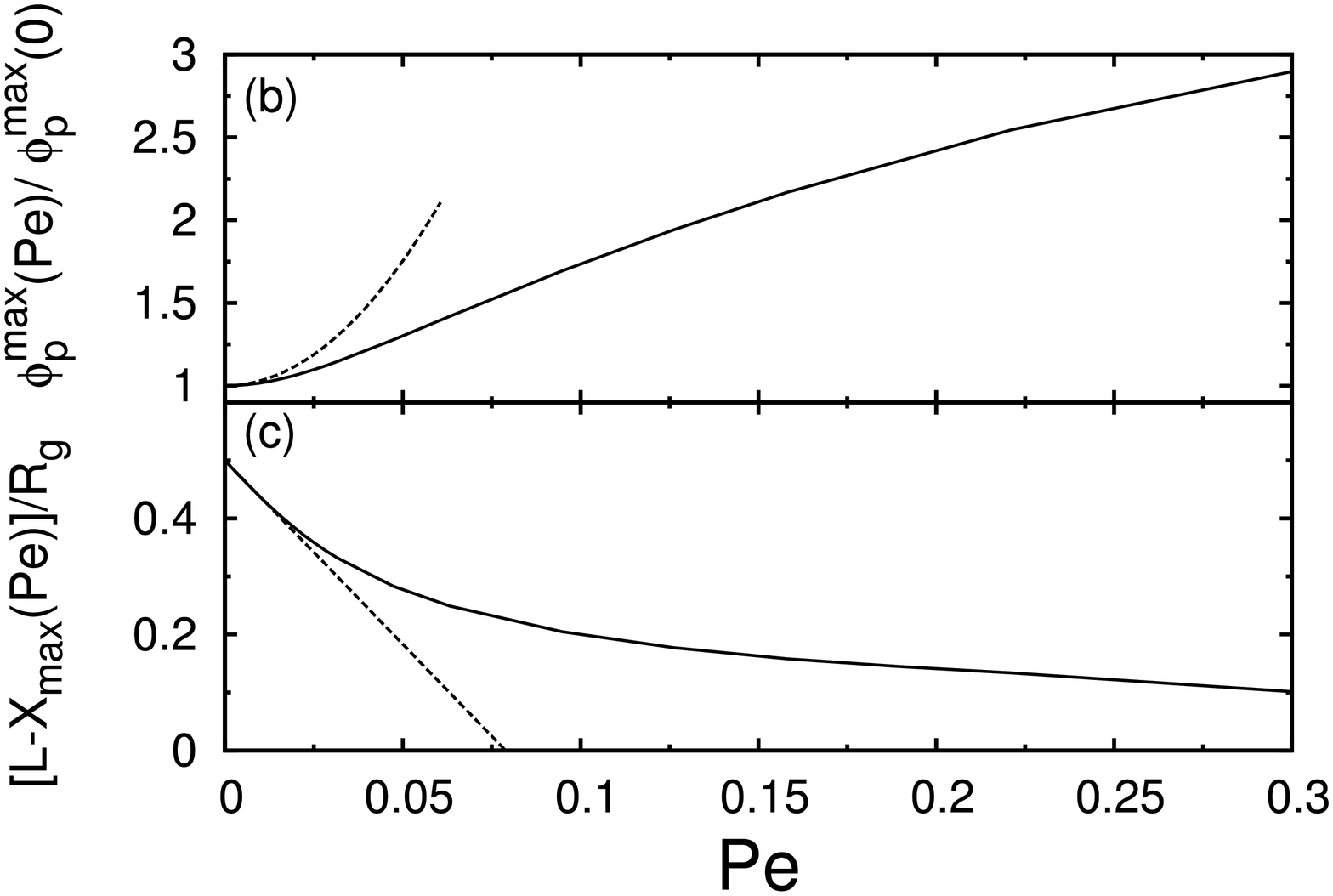}}}
\end{center}
\caption{
(a) Steady state volume fraction profiles 
    of polymer segments $\phi_\rp(x,\infty)$
    in the polymer solution 
    ($\bar \phi_\rp=0.1$, $N=1000$, $\chi=0.5$ and $L/R_\rg=1$)
    at \Pe= 0, 0.05, 0.1, 0.2, and 0.3.
(b) Peak polymer segment concentration $\phi_{\rp}^{\rm max}({\sf Pe})$ and 
(c) distance between the peak position $x_{\rm max}({\sf Pe})$
    and the downstream-side wall for various $\sf Pe$.
The continuous lines are numerical results 
from Eqs.(\ref{eqn:ChemicalPotential_2}) and (\ref{eqn:phi_2})
and the dotted lines are the 
analytical approximation given by 
Eqs.(\ref{eqn:peak_height}) and (\ref{eqn:peak_position}).
}
\label{fig:Fig_04_Profile_in_theta_solvent_large_Pe}
\end{figure}

To demonstrate how well the first-order approximation
describes the change of the concentration profile under a uniform flow, 
we consider the steady state case with $L/R_\rg=1$ 
under various Peclet numbers 
and compare the approximated profiles to the numerical results. 
Figure \ref{fig:Fig_02_graph_1d} shows 
the steady state concentration profiles of polymer segment 
under small Peclet number flows, {\em i.e.,} 
${\sf Pe} \le 1.5\times 10^{-2}$ 
in a polymer solution of 
$N=1000$, $\bar \phi_\rp=0.1$ and $\chi=0.5$. 
Note that $\bar\phi_\rp=0.1$ is slightly below the overlap volume fraction 
$\phi^\ast_\rp \simeq 0.11$ for polymer chains with $N=1000$.
The first-order analytical results are in good agreement with 
the numerical results 
under weak flow conditions ($\Pe\lsim 0.01$).

\begin{figure}
\begin{center}
\rotatebox{-90}{\resizebox{0.65\columnwidth}{!}{\includegraphics{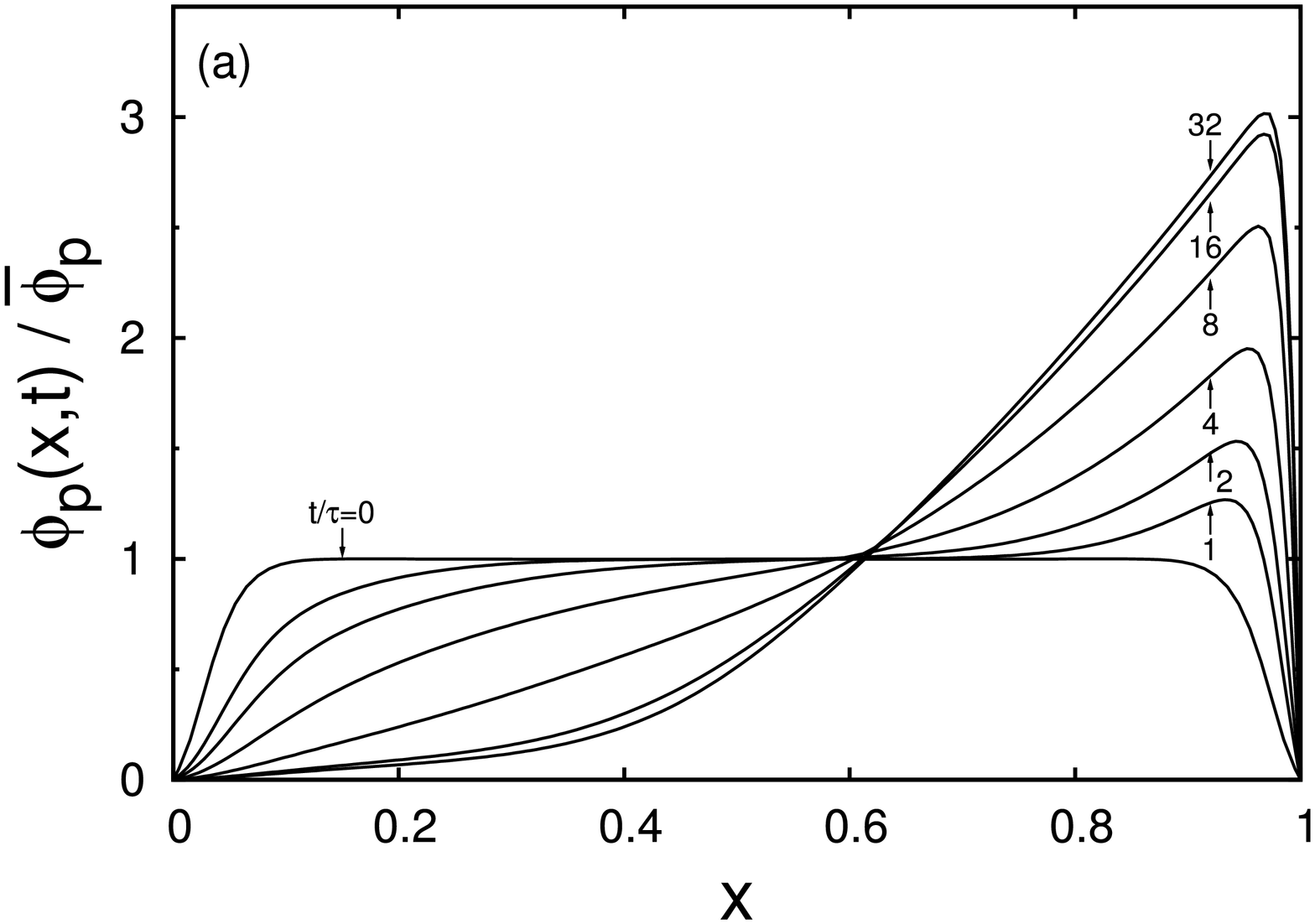}}}
\rotatebox{-90}{\resizebox{0.65\columnwidth}{!}{\includegraphics{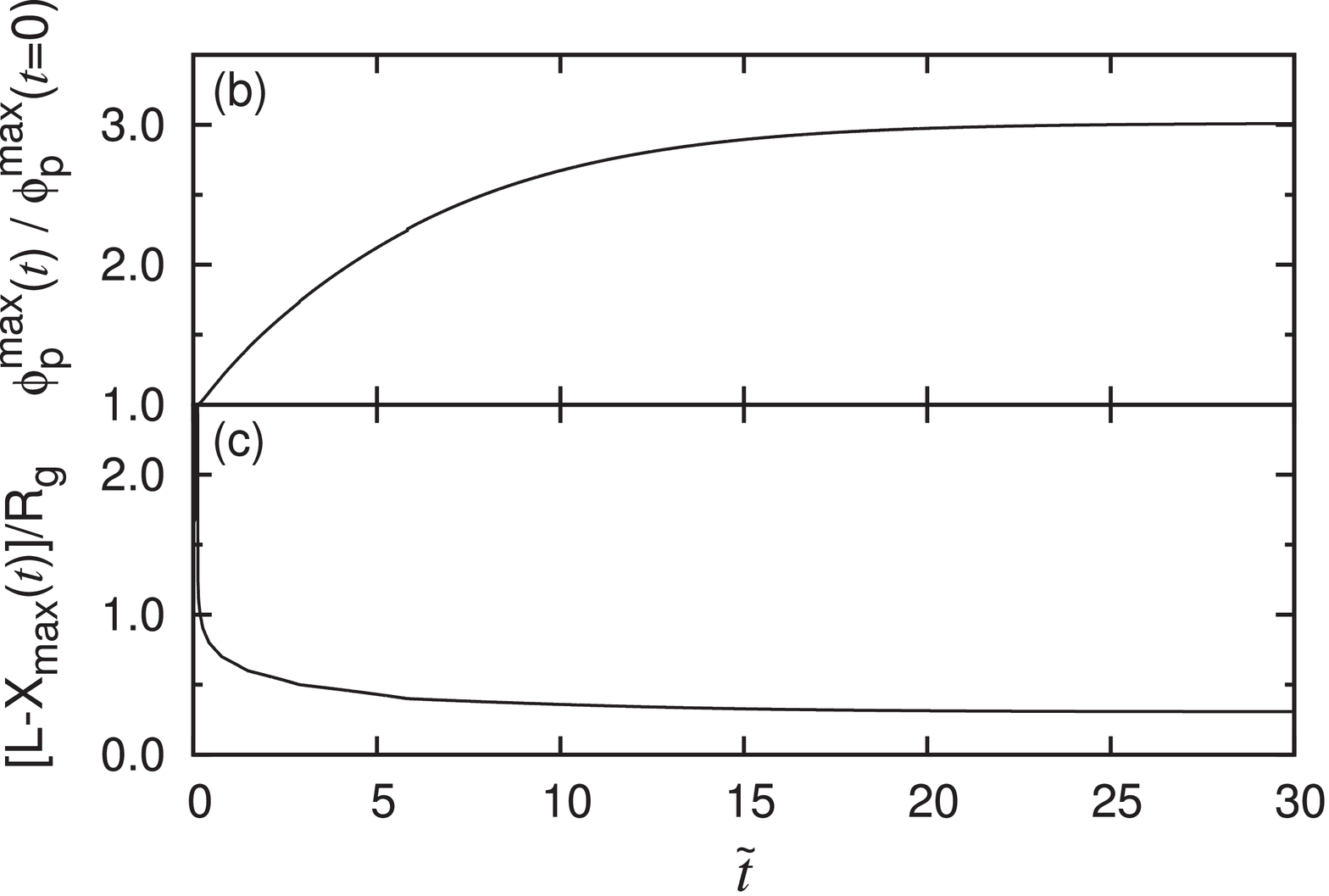}}}
\end{center}
\caption{
(a) Transient evolutions of
    the segment concentration profiles $\phi_\rp(x,t)$,  
(b) the peak concentration, and 
(c) the distance from the peak position 
    to the downstream-side wall.
Applied parameters are: 
$\bar \phi_\rp=0.1$, $N=1000$, $\chi=0.5$, 
$L/R_\rg=10$, and {\sf Pe}=0.05 at $\tilde t=t/\tau \ge 0$.
The profiles are at $\tilde t=$0 (no flow), 
1, 2, 4, 8, 16, and 32. 
At $\tt \simeq $20, 
the profile reaches a steady state as seen from (b) and (c). 
Note that the results shown in 
Fig.\ref{fig:Fig_03_profile_time_evo_Pe=0.1_chi=0.5_phi0=0.01_Lx=1} 
are based on GSA, 
here the profiles are calculated by using the dynamic SCFT scheme.
}
\label{fig:Fig_05_profile_time_evo_Pe=0.1_chi=0.5_phi0=0.01_Lx=10_DynSCF}
\end{figure}
%
%
%
%
\begin{figure}
\begin{center}
\rotatebox{-90}{\resizebox{0.65\columnwidth}{!}{\includegraphics{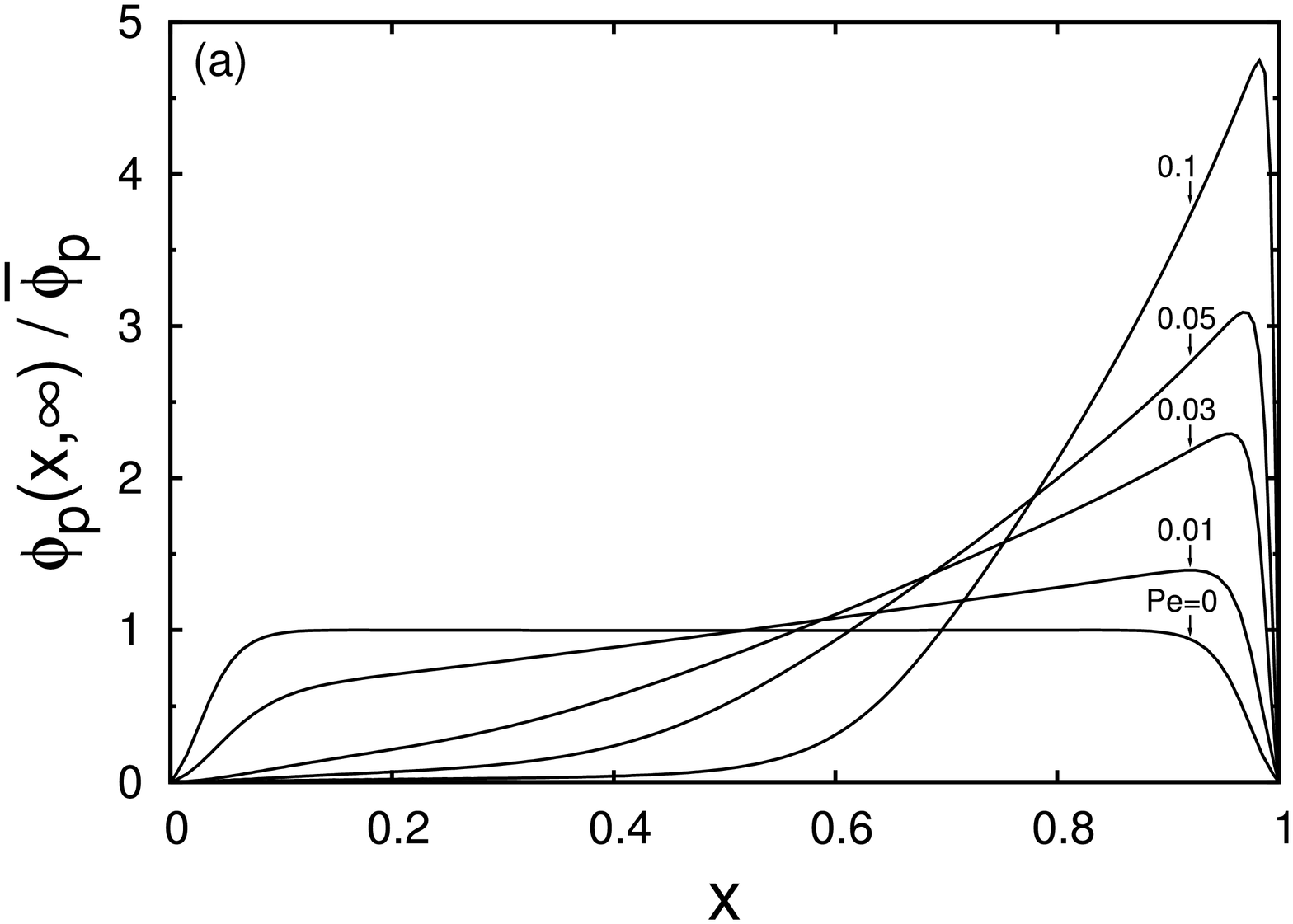}}}
\rotatebox{-90}{\resizebox{0.60\columnwidth}{!}{\includegraphics{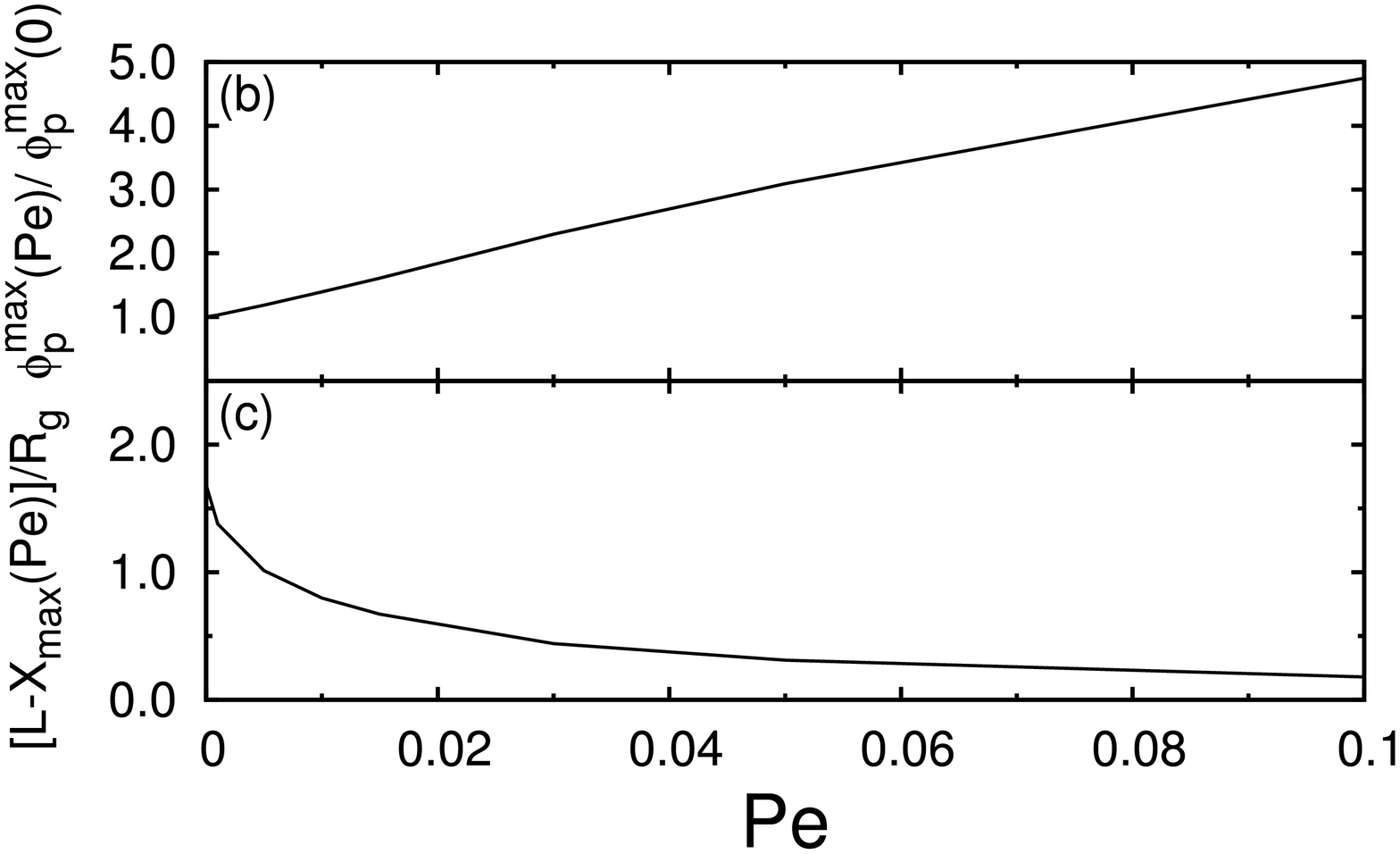}}}
\end{center}
\caption{
(a) Steady state volume fraction profiles of polymer segments 
    in the polymer solution 
    ($\bar \phi_\rp=0.1$, $N=1000$, $\chi=0.5$, $L/R_\rg=10$) at 
    \Pe = 0, 0.01, 0.03, 0.05 and 0.1.
    (b, c) Change of 
    the peak concentration $\phi_{\rp}^{\rm max}({\sf Pe})$ and 
    the distance from the peak position $x_{\rm max}({\sf Pe})$
    to the downstream-side wall for various $\sf Pe$.
    The data are computed by the dynamic SCFT scheme. 
}
\label{fig:Fig_06_Profile_in_theta_solvent_Lx=10}
\end{figure}
%
%
Figure \ref{fig:Fig_03_profile_time_evo_Pe=0.1_chi=0.5_phi0=0.01_Lx=1}
shows the transient behavior of $\phi_\rp(x,t)$ 
in a slit with $L/R_\rg=1$ 
after imposing a fluid flow with 
\Pe=0.05 to an equilibrium profile at $t=0$.
The time evolutions of the peak value $\phi_{\rp}^{\rm max}(t)$
and 
the distance from the position $x_{\rm max}(t)$ to the downstream-side wall
are shown 
in Fig.\ref{fig:Fig_03_profile_time_evo_Pe=0.1_chi=0.5_phi0=0.01_Lx=1}(b) 
and (c),
respectively.
The shift of the peak position to the steady state position 
is slightly faster than that of the peak concentration, {\em i.e.,} 
the peak position reaches steady state first,  
and then the concentration profile 
becomes sharper gradually.
This indicates that a strong convective effect applies to 
the polymer segments 
and relatively slow relaxation of 
the polymer distribution across the slit.
The time the system needs to reach the steady state
is approximately 15$\tau$ to 20$\tau$ for $\Pe=0.05$, 
which is consistent with the accumulation time. 
In Fig.\ref{fig:Fig_04_Profile_in_theta_solvent_large_Pe}, 
we show that the steady state concentration profiles $\phi_\rp(x,\infty)$
evolve upon increasing the flow rate characterized by 
{\sf Pe}=0, 0.05, 0.1, 0.2 and 0.3 in a polymer solution with 
$\bar \phi_\rp=0.1$, $N=1000$ and $\chi=0.5$.
Figure \ref{fig:Fig_04_Profile_in_theta_solvent_large_Pe}(b) and (c) show
the peak height $\phi_{\rp}^{\rm max}({\sf Pe})$ of the concentration profile 
and 
the distance from the peak position $x_{\rm max}({\sf Pe})$
to the downstream-side wall
at steady state under various flow strengths.
The peak height $\phi_{\rp}^{\rm max}$ increases quadratically with {\sf Pe} 
for small $\sf Pe$ as expected in Eq.(\ref{eqn:peak_height}).
In contrast to the peak height $\phi_{\rp}^{\rm max}$, 
the shift of the peak position for 
${\sf Pe} \lsim 0.02$  is well described by 
a linear approximation of Eq.(\ref{eqn:peak_position}) 
as seen from Fig.\ref{fig:Fig_04_Profile_in_theta_solvent_large_Pe}(c).
For ${\sf Pe} \gsim 0.02$, 
the shift of peak position to the downstream side is suppressed
due to the wall effect. 
Specifically from Eq.(\ref{eqn:phi_2}),
the peak position is determined by the competition 
of polymer segment fluxes induced by 
the hydrodynamic flow and by the thermodynamic force. 
The suppression of the peak shift distance is due to 
the increase of the thermodynamic flux 
by accumulating polymer chains at the downstream side. 
As seen in Fig.\ref{fig:Fig_04_Profile_in_theta_solvent_large_Pe}(b)
(and later in Fig.6(b), Fig.8(c) and Fig.10(b)),  
the segment volume fraction at the peak 
seems to increase indefinitely with Pe, 
while the peak position seems to saturate. 
This is because we restrict our study to a dilute to 
semi-dilute polymer solution, 
and we have omitted the $\phi_\rs^2$-factor 
in the effective diffusion coefficient 
of eq.(\ref{eqn:phi_2}). 
This factor appears in the original transport equation, 
Eq.(\ref{eqn:eq_of_continuity2}). 
If the volume fraction of polymer segment at the peak
comes close to unity, $\phi_\rs^2 \rightarrow 0$, 
the effective diffusivity approaches zero around the peak. 
In such a case, however, 
the dilute to semi-dilute assumption is no longer valid, 
and the non-uniform velocity and 
the viscoelastic effects should be taken into account.
This complicated scenario will be investigated 
in future work.
%
%
\begin{figure}[t]
\begin{center}
\rotatebox{-90}{\resizebox{0.70\columnwidth}{!}{\includegraphics{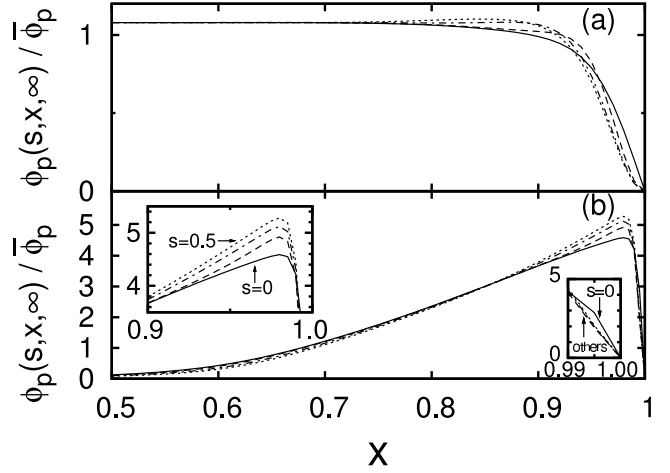}}}
\end{center}
\caption{
Concentration profiles of $s$-th segment 
in the polymer solution ($\bar \phi_\rp=0.1$, $N=1000$, $\chi=0.5$, 
$L/R_\rg=10$)
at steady states under 
(a) {\sf Pe}=0 (no flow) and 
(b) {\sf Pe}=0.1 in the region $0.5\le x \le 1$. 
The lines are for 
$s=0$ (solid line), 0.05 (dashed), 0.2 (dash dotted) and 0.5 (dotted).
The two insets in (b) 
show segment profiles very close to the downstream-side wall. 
}
\label{fig:Fig_07_profile_of_s-th_seg_Pe=0.1_chi=0.5_Lx=10}
\end{figure}

\subsubsection{Wide Slit, $L > \pi R_\rg$}

When the distance $L$ is larger than $\pi R_\rg$, 
GSA can not be used 
to evaluate the chemical potential difference $\mu(\br)$.
Here we utilize the scheme (i)-(iv) based on DSCFT
explained in Sec.~\ref{subsec:Chemical Potential of Polymers}.
To demonstrate the scheme, 
we consider the case of $L/R_\rg$=10. 
Firstly, we show the time evolution of the concentration profiles 
under \Pe=0.05
in Fig.\ref{fig:Fig_05_profile_time_evo_Pe=0.1_chi=0.5_phi0=0.01_Lx=10_DynSCF},
in which 
the volume fraction profile of the polymer solution ($\chi=0.5$) 
at a quiescent state ($\sf Pe$=0) 
coincides with the analytical result of the depletion profile 
near a single wall \cite{RemcoBook2011,Taniguchi1992,Eisenriegler1996}. 
As $t\rightarrow \infty$, 
the peak appears near the end of the depletion zone 
along the downstream side.
In comparison with the narrow slit case, 
$x_{\rm max}(t)$ shifts to the steady state position 
faster than that of the peak height 
(Fig.\ref{fig:Fig_05_profile_time_evo_Pe=0.1_chi=0.5_phi0=0.01_Lx=10_DynSCF}(b) 
and (c)). 
Although this tendency has been found in the narrow slit case 
(Fig.\ref{fig:Fig_03_profile_time_evo_Pe=0.1_chi=0.5_phi0=0.01_Lx=1}),
it is enhanced in the wider slit case as seen in 
Fig.\ref{fig:Fig_05_profile_time_evo_Pe=0.1_chi=0.5_phi0=0.01_Lx=10_DynSCF}.
The time the system needs to reach steady state 
is approximately $20\tau$ as seen from 
Fig.\ref{fig:Fig_05_profile_time_evo_Pe=0.1_chi=0.5_phi0=0.01_Lx=10_DynSCF}(a)
and (b).
The accumulation time ($t^\ast\sim 20\tau$) gives a 
better estimation for $\Pe=0.05$ than the narrow slit case 
described in sec.\ref{sec:Narrow_slit_theta_solvent}.
This good agreement 
is owing to the good estimation of the transport distance 
of the polymer segments $L$ in the accumulation time.

Figure \ref{fig:Fig_06_Profile_in_theta_solvent_Lx=10} 
shows the steady state concentration profiles 
for slit width $L/R_\rg=10$,  
\Pe=0 to 0.1 in the polymer solution of 
$\bar \phi_\rp=0.1$, $N=1000$ and $\chi=0.5$.
These are numerical results obtained by 
the dynamic SCFT scheme described in Eq.(\ref{eqn:eq_of_continuity2}) 
and Eqs.(\ref{eq:Helmholtz2})-(\ref{eq:mu_s}). 
In addition, 
the peak concentration and its position 
at steady state are functions of $\sf Pe$.  
For $\sf Pe \gsim$ 0.05 
the peak position does not change much,  
but the peak height continues to increase with $\sf Pe$.
This indicates that the segment accumulation becomes narrower 
while keeping the peak position almost the same as {\sf Pe} increases,
which is due to the a strong depletion effect from the wall.
%
%
\begin{figure}
\begin{center}
\rotatebox{-90}{\resizebox{0.65\columnwidth}{!}{\includegraphics{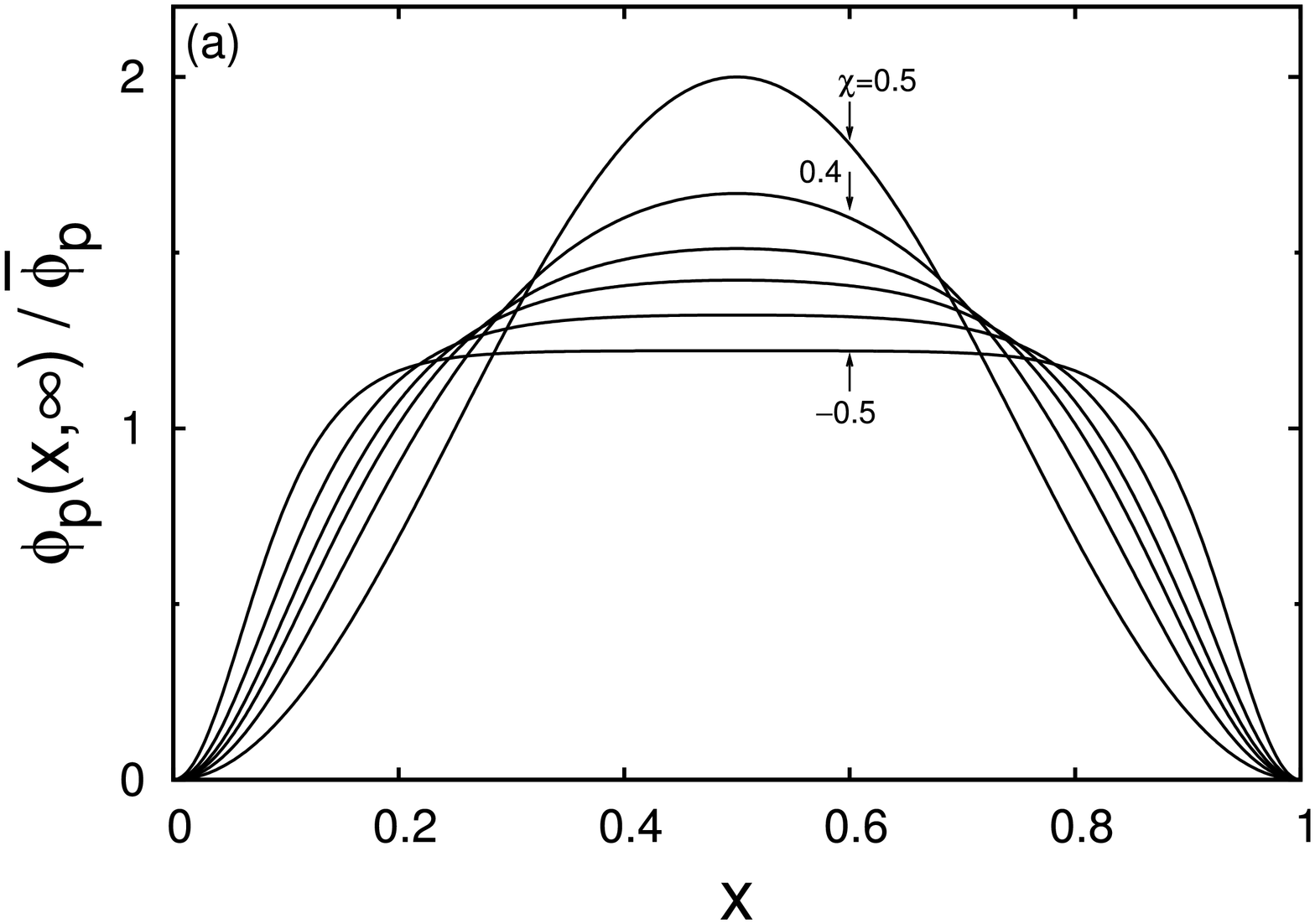}}}
\rotatebox{-90}{\resizebox{0.65\columnwidth}{!}{\includegraphics{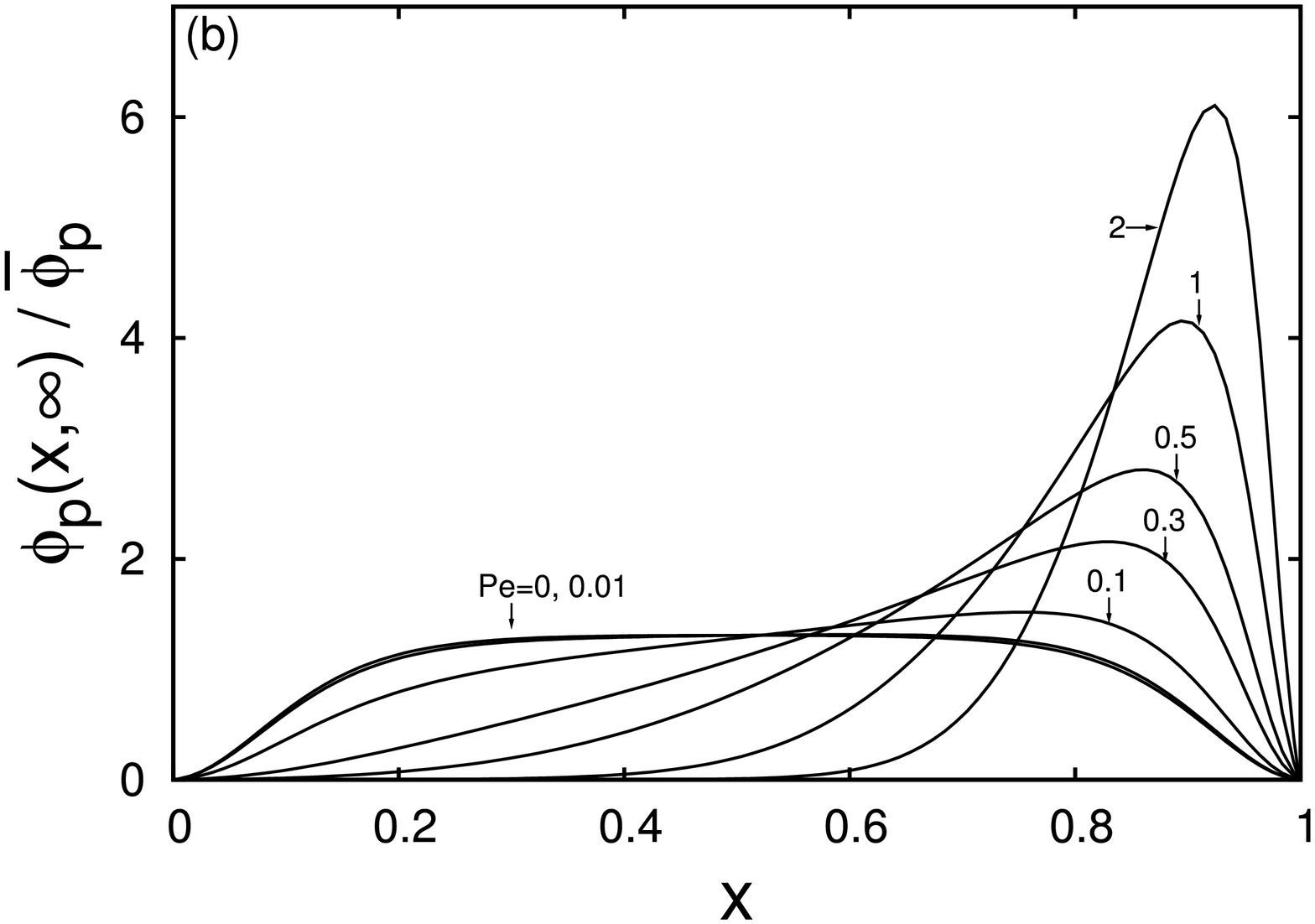}}}
\rotatebox{-90}{\resizebox{0.65\columnwidth}{!}{\includegraphics{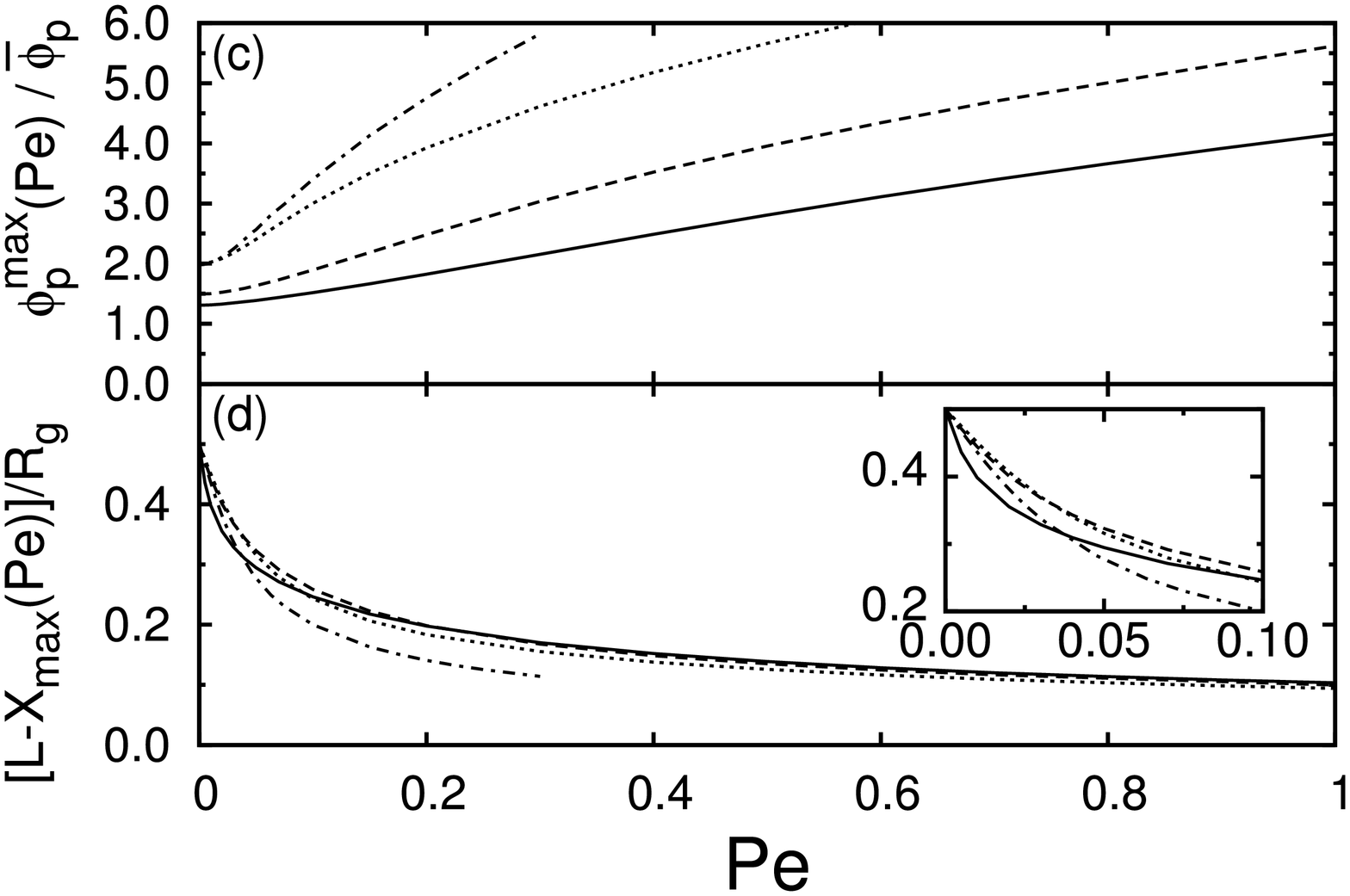}}}
\end{center}
\caption{
\baselineskip=12pt 
Steady state results for good solvent conditions
in a narrow slit ($L/R_\rg=1$, $\bar \phi_\rp=0.1$, $N=1000$).  
(a) Equilibrium ({\sf Pe}=0) concentration profiles of polymer segments 
    in good solvents with $\chi=-$0.5, 0.0, 0.2, 0.3, 0.4, 
    and in $\Theta$-solvent ($\chi=0.5$)
    from the bottom to the top lines near the middle point.
(b) The polymer segment volume fraction profiles 
for $\chi=0$ and {\sf Pe}=0 to 2. 
The peak height (c) and 
the distance from the peak position to the downstream-side wall (d) 
for $\chi$=0 (solid line), 
and 0.3 (dashed line) 
are compared with the $\Theta$-solvent 
($\chi=0.5$, $m=1$, dash-dotted line) results.
As a reference, 
the test case of $\chi=0.5$ and $m=3/4$ is 
shown by the dotted line. 
The graph of the $\Theta$-solvent is truncated at 
a peak height higher than 6$\bar \phi_\rp$. 
}
\label{fig:Fig_08_steady_state_chi=0.0_Lx=1.eps}
\end{figure}
%
%
\begin{figure}
\begin{center}
\rotatebox{-90}{\resizebox{0.63\columnwidth}{!}{\includegraphics{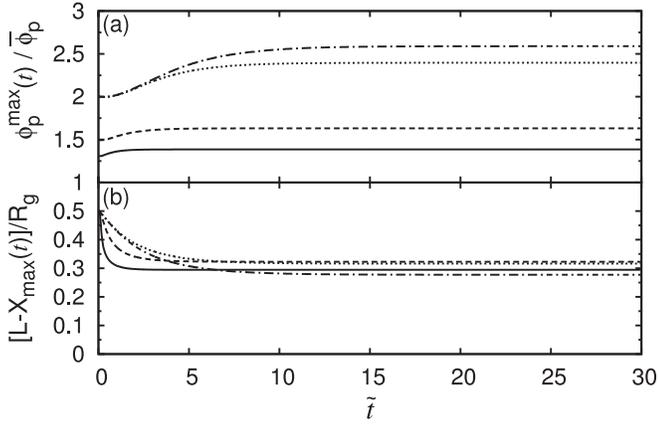}}}
\end{center}
\caption{
The time evolution of 
the peak concentration (a) and 
distance (b) from the peak position to the downstream-side wall
in polymer solutions 
within a narrow slit ($L/R_\rg=1$, $\bar \phi_\rp=0.1$, $N=1000$) 
under steady flows with $\sf Pe=0.05$ 
for
$\chi$=0 (solid line),
0.3 (dashed line), and
0.5 ($\Theta$-solvent, $m=1$, dash-dotted line). 
As a reference, 
the case of $\chi=0.5$, but with $m=3/4$ is shown by the dotted line. 
}
\label{fig:Fig_09.eps}
\end{figure}

The individual segment concentration provides further details 
of the convective effect.
The statistical weight $q(x,s)$ obtained by Eq.(\ref{eqn:q})
through the procedures (i)-(iv) 
is used to calculate the concentration profile $\phi_\rp(s,x,t)$
for the $s$-th segment :
\begin{equation}
\phi_\rp(s,x,t)= { V \bar \phi_\rp \over Q_\rp } q(s,x)q(1-s,x).
\end{equation}
%
%
\noindent
Figure \ref{fig:Fig_07_profile_of_s-th_seg_Pe=0.1_chi=0.5_Lx=10} 
shows the concentration profiles for the $s$-th segment 
at steady state for $s=$0.0 (end point), 0.05, 0.20, and 0.5 (midpoint)
under two conditions: 
(a) {\sf Pe}=0 (no flow) and 
(b) {\sf Pe}=0.1. 
Only the region 
$0.5 \le x \le 1$ is presented.
The profile for {\sf Pe}=0 is symmetric about 
$x$=0.5, and for \Pe=0.1
the concentration is very low for $0 \le x \le 0.5$.
The distortion of the polymer segment concentration   
is significantly enhanced by the flow. 
Furthermore, in the case of {\sf Pe}=0.1, 
the distribution of the end segment ($s=0$) is the broadest 
among others and the distributions for $0.1 \lsim s < 0.5$
are similar to the case of $s=0.5$ (midpoint) 
as expected \cite{Eisenriegler1983,Eisenriegler1993,Eisenriegler2002}. 
%
%
%
%

\subsection{Good-solvent cases}

\subsubsection{Narrow Slit, $L < \pi R_\rg$}

For good-solvent condition ($\chi < 1/2$, $m =3/4$) under steady flow, 
the volume fraction profile can be obtained 
from Eq.(\ref{eqn:phi_steady_state}). 
When {\sf Pe}=0, 
we obtain the analytical volume fraction profile as
(see Fig.\ref{fig:Fig_08_steady_state_chi=0.0_Lx=1.eps}(a), 
derivation is given in Appendix \ref{sec:Appdx:Good_Sovent})
\begin{equation}
 \phi_\rp(x) = \varphi_\ro^2(x) 
            = [\varphi_{\rm m} \sn(\tx, k)]^2, 
\label{eqn:varphi_elliptic} 
\end{equation}
where 
$\varphi_{\rm m}$ is the maximum value of $\sqrt{\phi_\rp(x)}$, 
$\sn(\tilde x,k)$ is the Jacobi elliptic integral, 
$\tilde x= \varphi_\Rm x \sqrt{v/2k^2} L/\ell$, 
and the constant $k$ is defined by 
$k^2 = {v \varphi_\Rm^2 /[ 2\mu_\ro(0)  - v \varphi_\Rm^2]}.$
From Eq.(\ref{eqn:first_order_equation})
we obtain the first-order asymptotic equation 
under a weak flow condition (${\sf Pe} \ll 1$):
\begin{eqnarray}
&& 
\biggr [~  {d^2  \over d\tx^2}
             + k^2 + 1 - 6k^2 \sn^2(\tx,k)
  ~\biggr ] \tilde \varphi_1(\tx) 
= 
\nonumber
\\ 
&& 
- {L^2 \over 4 \ell^2 K^2(k) }\sn(\tx,k)
  \biggr [   \mu_1(0)  
            +  {\varphi_\Rm \over 2K(k)} 
                          \int_0^\tx \sn({\tilde x}',k) d{\tilde x}'
  \biggr ], 
\quad\quad
\label{eqn:Appdx:first_order_equation_2}
\end{eqnarray}
where $\tilde \varphi_1=\varphi_1/\varphi_\Rm$.
The constant $\mu_1(0)$ is determined by 
the normalization condition 
$\int_0^{1} \varphi^2(x) dx = 1$. 
Unlike the $\Theta$-solvent case, 
analytic approximations are extremely involved here, 
and thus the concentration profiles are 
obtained numerically for both weak and strong flow conditions. 
%
%
\begin{figure}
\begin{center}
\rotatebox{-90}{\resizebox{0.65\columnwidth}{!}{\includegraphics{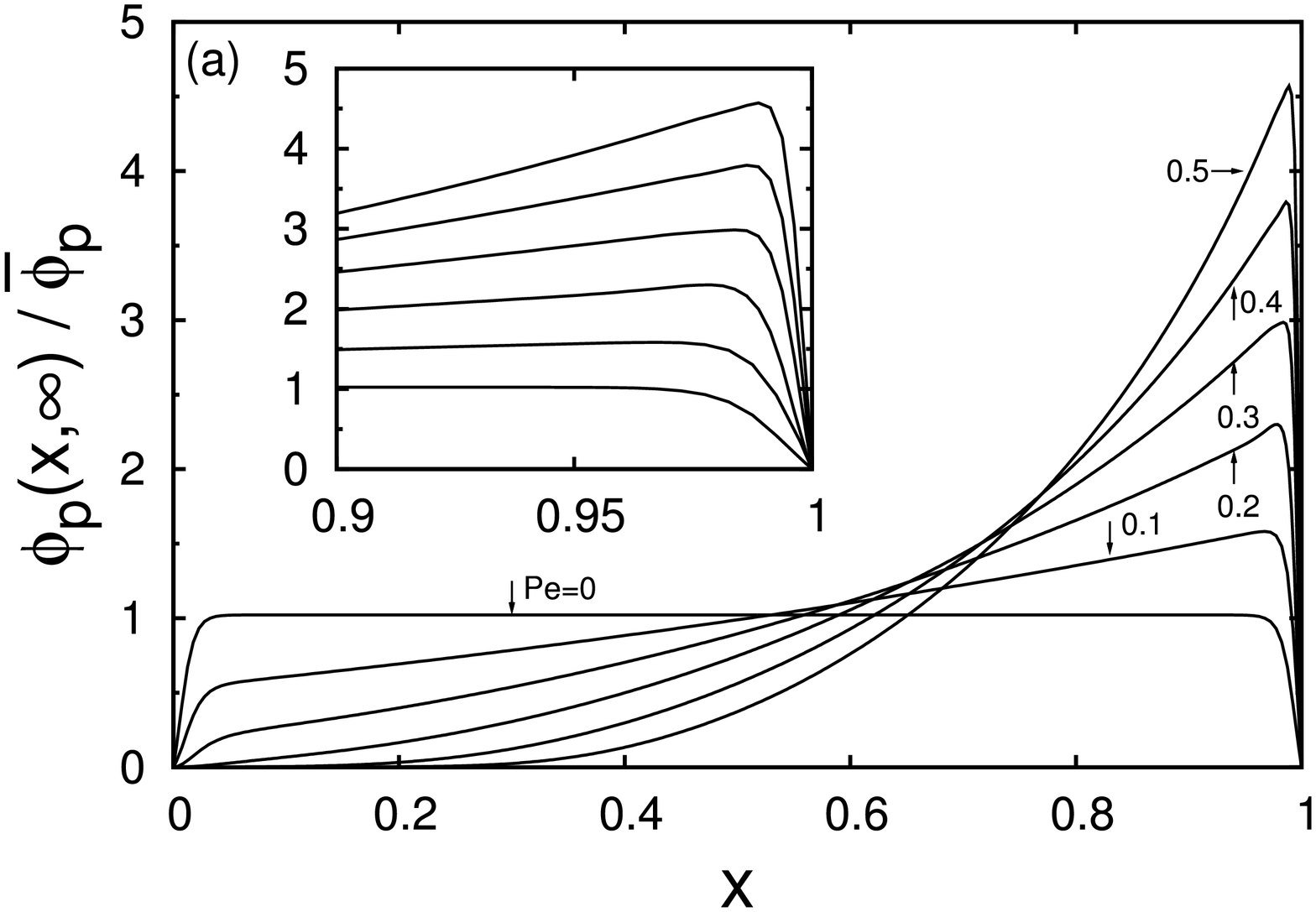}}}
\rotatebox{-90}{\resizebox{0.65\columnwidth}{!}{\includegraphics{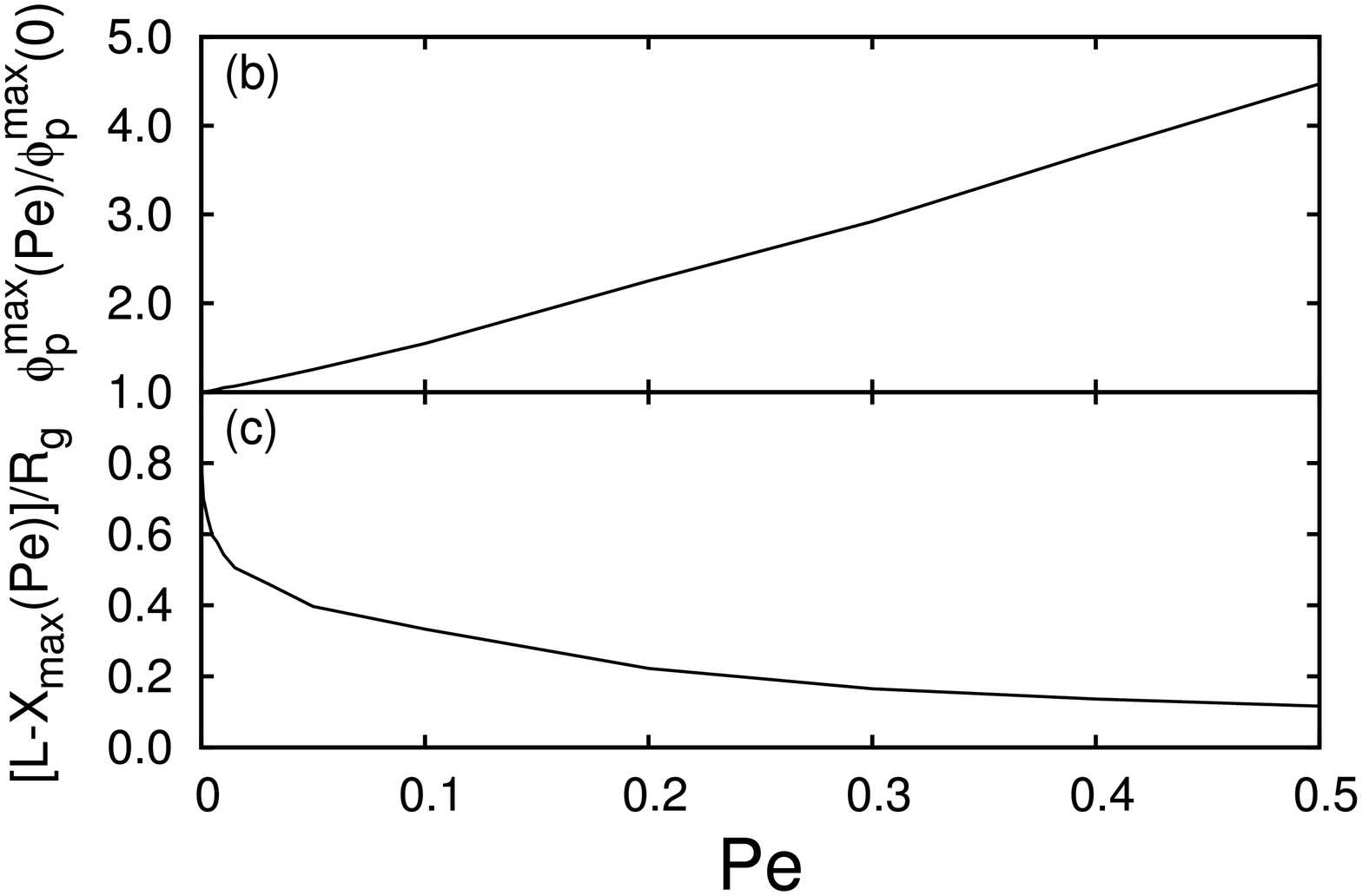}}}
\end{center}
\caption{
(a) Steady state profiles
    of a polymer solution 
    ($\bar \phi_\rp=0.1$, $N=1000$, and $\chi=0$)
    in a wide slit with $L/R_\rg=10$ 
    under steady flows with 
    \Pe=0, to 0.5, 
(b) the peak height $\phi_{\rp}^{\rm max}({\sf Pe})$,  
    and 
(c) the distance from the peak position 
    to the downstream-side wall
    as a function of ${\sf Pe}$. 
The inset in (a) shows the result near the wall.
The results are computed by using the dynamical SCFT scheme. 
}
\label{fig:Fig_10_steady_state_chi=0.0_Lx=10.eps}
\end{figure}
%
%
\begin{figure}
\begin{center}
\rotatebox{-90}{\resizebox{0.640\columnwidth}{!}{\includegraphics{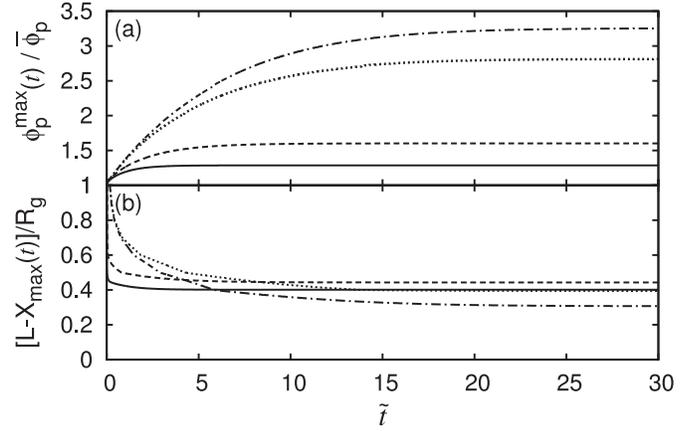}}}
\end{center}
\caption{
The time evolution of 
the peak concentration (a) and 
distance (b) from the peak position to the downstream-side wall
in polymer solutions 
within a wide slit ($L/R_\rg=10$, $\bar \phi_\rp=0.1$, $N=1000$) 
under steady flows with $\sf Pe=0.05$ 
for $\chi$=0 (solid line), 0.3 (dashed line), 
and 0.5 ($\Theta$-solvent, $\chi=0.5$, $m=1$, dash-dotted line). 
As a reference, 
the case of $\chi=0.5$, but with $m=3/4$ is shown by the dotted line. 
}
\label{fig:Fig_11}
\end{figure}
%
%
In Fig.\ref{fig:Fig_08_steady_state_chi=0.0_Lx=1.eps}(b),
we plot steady state volume fraction profiles 
$\phi_\rp(x)$ of a polymer solution 
for various {\sf Pe}-values. 
In Fig.\ref{fig:Fig_08_steady_state_chi=0.0_Lx=1.eps}(c) and (d), 
both the peak height and position are plotted 
as functions of {\sf Pe} for various $\chi$-values.
As a reference,  
the $\Theta$-solvent ($\chi=0.5$, $m=1$) result given 
in Fig.\ref{fig:Fig_04_Profile_in_theta_solvent_large_Pe} 
is shown by the dash-dotted lines, 
and the case of $\chi=0.5$ and $m=3/4$ 
is shown by the dotted line.
From the comparison among the cases of $\chi=0$, 0.3, and 0.5 with $m$=3/4
in Fig.\ref{fig:Fig_08_steady_state_chi=0.0_Lx=1.eps}(c), 
we find that the peak height of 
polymers concentration in a good solvent is less-sensitive 
to the convective effect with lower $\chi$.
This is caused by the enhanced excluded volume effect for better solvency.
On the other hand, 
the behaviors of the peak height in the $\Theta$-solvent 
are slightly different from those in good-solvent conditions. 
The different behavior between the good- and $\Theta$-solvents
comes from 
the $\phi_\rp$-dependent diffusion coefficient in Eq.(\ref{eqn:phi_2}).
This can be confirmed from the evidence that 
the peak position for $\chi=0.5$ but with $m=3/4$ 
exhibits a similar $\Pe$-dependence 
to those in good solvents where $m=3/4$.

Figure \ref{fig:Fig_09.eps} shows that  
a polymer solution with smaller $\chi$ 
exhibits a smaller peak height 
and reaches steady state faster. 
When $\chi=0$, only a small change of peak height 
appears due to the larger excluded volume effect 
under a better solvency.
However, 
the distance between the peak position and the wall 
on the downstream side
for $\chi=$0.3 is larger than those in $\chi=0.0$ and $\chi=0.5$. 
The reason is because that 
the depletion thickness in $\chi=0$ 
is smaller than those of $\chi=0.3$ in a quiescent state \cite{Fleer2003},  
and therefore under a flow,  
initially the peak is formed at a position relatively 
closer to the wall for $\chi=0$.

Although the depletion thickness at a quiescent state
for $\chi=0.5$ is larger than for $\chi=0.3$, 
the peak can easily be shifted by the flow 
compared to the two good solvent cases, 
and therefore the peak position becomes closer to the wall 
for $\chi=0.5$ than for the $\chi=0.3$ case. 
As seen in Fig.\ref{fig:Fig_09.eps}, 
under a good solvent condition, 
the time for systems to reach the steady state under $\Pe=0.05$ 
is approximately $2\tau$ for $\chi=0$ and $3\tau$ for $\chi=0.3$, 
which are much shorter than 
the estimated accumulation time ($t^\ast \sim 20\tau$) 
due to the excluded volume effect.

\subsubsection{Wide Slit, $L/R_\rg=10$}

Figure \ref{fig:Fig_10_steady_state_chi=0.0_Lx=10.eps} 
shows the results for the same polymer solution 
in a wide slit with $L/R_\rg=10$ and $0 \le {\sf Pe} \le 0.5$. 
The steady state profiles of polymer segment volume fraction 
under various $\sf Pe$ are shown in (a). 
The peak concentration and position are plotted in Figures (b) and (c).
As seen from (a) and (c),  
the profile is almost flat in the quiescent state and 
by applying a flow 
the peak position jumps from the center of the slit 
to a downstream position near the wall.
The polymer segment profile is strongly influenced by the applied flow. 
Because the depletion thickness under good solvent conditions is 
relatively small compared to the slit width, 
the profile has a sharp peak.  
This is also because the depletion thickness is reduced 
due to good solvency or large excluded volume effect 
among the polymer chains. 

Figure \ref{fig:Fig_11} shows that 
the concentration profiles in good solvents reach steady states 
much faster than in a $\Theta$-solvent, 
similar to the good solvent case shown in Fig.\ref{fig:Fig_09.eps}.
The shorter accumulation time comes from 
a smaller depletion thickness 
for a better solvency at the quiescent state,
such that the peak appears at the very end of depletion zone.
As a result, the accumulation time is much smaller than
the estimated one.

\begin{figure}
\begin{center}
\rotatebox{-90}{\resizebox{0.67\columnwidth}{!}{\includegraphics{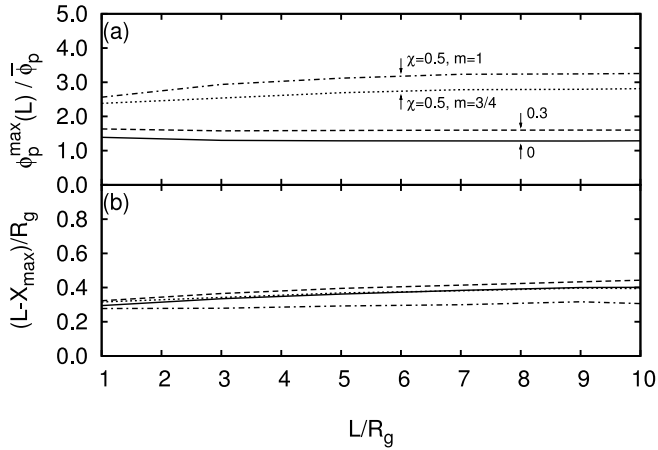}}}
\end{center}
\caption{
Effect of slit width.
(a) The peak volume fraction and 
(b) the distance from the peak position to the wall in the downstream side 
in polymer solutions ($\bar \phi_\rp=0.1$, $N=1000$) 
at steady states under a flow with $\sf Pe=0.05$  
for 
$\chi$=0 (solid line), 
0.3 (dashed line), 
the $\Theta$-solvent case ($\chi=0.5, m=1$) (dash-dotted line), 
and the case with $\chi=0.5$ and  $m=3/4$   (dotted line).
}
\label{fig:Fig_12.eps}
\end{figure}
In Fig.\ref{fig:Fig_12.eps}
we demonstrate the dependency of the slit width for $\Pe=0.05$ 
under solvent conditions $\chi=$0.0, 0.3 and 0.5.
The peak concentrations $\phi_\rp^{\rm max}(L)$ 
for the given $\chi$-values 
are not sensitive to $L/R_\rg$.
In a $\Theta$-solvent ($\chi=0.5$, $m$=1) and 
the test case with $\chi=0.5$ and $m=3/4$, 
the peak concentrations increase somewhat with $L$.
In good solvents, however, 
the peaks slightly decrease with increasing $L$. 
Under a flow condition, 
the polymer segment accumulation is determined by the competition between
the hydrodynamic flux and the diffusive thermodynamic flux.
The thermodynamic flux is contributed by 
the excluded volume effect 
and the translational entropy that suppresses 
the polymer segment accumulation 
at the downstream side. 
The increase of the peak concentration in a $\Theta$-solvent 
might come from rather weak excluded volume effect. 
In good solvents, the excluded volume effect lowers 
the peak height and the translational entropy 
also suppresses the accumulation 
when the slit width becomes large ($L \gg R_\rg$). 
As seen from Fig.\ref{fig:Fig_12.eps}(b), 
the distance between the peak position and the downstream-side wall 
slightly increases with $L$ for all the $\chi$-values
under a fixed $\Pe$. 
In the range of $1 \le L/R_\rg \le$10,
the distance is roughly less than $0.1R_\rg$ for $\chi=$0.0, 0.3 and 0.5,
which means that 
the depletion thickness 
in a flow field
is mainly determined by $\Pe$. 
Irrespective of the slit width, 
a better solvency always exhibits
a weaker polymer segment accumulation 
because of the larger exclude volume effect. 
On the other hand, 
the shift distance is larger for $\chi=$0.3 
than for the $\chi=0.0$ and $\chi=0.5$ cases
due to the same reason explained 
when discussing Fig.\ref{fig:Fig_09.eps}.

Finally, because we used DSCFT scheme 
to obtain steady state results for $L/R_\rg>1$, 
the computational expense is much higher 
than for the $L/R_\rg=1$ case using GSA, 
which makes it difficult to 
reach sufficient accuracy for $L\gg R_\rg$.
The distance between the peak position and the wall 
at the steady state 
slightly increases with $L$ as seen in Fig.\ref{fig:Fig_12.eps}(b), 
we conclude that the distance is almost constant against $L$ 
and there is no clear physical reason why 
the distance should increase with $L$ for fixed $\sf Pe$.
%
%


\section{Conclusions}
\label{sec:Conclusions}

The two-fluid model and the dynamic self-consistent field theory are 
successfully combined to characterize the polymer segment dynamics 
under the influence of a uniform flow and polymer depletion effect. 
This conceptual model demonstrates 
the segment concentration profile of a polymer solution ($0 \le \chi \le 0.5$) 
confined in between two parallel and solvent permeable walls. 
The continuous concentration profiles in transient and steady state analysis 
are characterized by the Peclet number \Pe, excluded volume parameter $\nu$, 
and the slit width $L/R_\rg$. 
The polymer segments accumulate at the downstream due to the convective effect. 
The competition between the hydrodynamic flux 
and the diffusive thermodynamic flux 
are featured by the height and position of the concentration peak. 
The mean flow transports polymer segments to the downstream and 
the thermodynamic flux acts to minimize the concentration gradient in bulk
and imposes a depletion region near the walls to suppress 
the loss of conformation entropy of the polymer chains. 
We provide analytical ground state approximation of the concentration profiles 
for the steady state and narrow slit case 
under a weak flow with theta or good solvents. 
All wide slit cases in either weak or strong flow are resolved numerically 
using the dynamic self-consistent field theory 
to distinguish the individual segment-level profiles influenced by the flow.

Using DSCFT, we find that the distribution of the end segment is the broadest 
as compared to other segments that tend 
to follow the distribution of the center-of-mass segments. 
At steady state, regardless the slit width, 
the peak concentration in a good solvent is less sensitive to 
the flow effect with decreasing $\chi$. 
This is due to the strong excluded volume effect in the good solvent.
Such behavior does not appear in the theta-solvent case 
because the $\phi_\rp$-dependent diffusion coefficient vanishes. 
The peak concentration and its location are mainly determined 
by the flow strength and the excluded volume effect, 
and weakly depend on the slit width. 
In transient analysis, we characterize the accumulation time 
for both narrow and wide slits. 
Regardless the solvent condition and the slit width, 
the peak location responses faster and 
reaches steady state earlier than the peak height. 
The estimated accumulation time is best applicable for the theta solvent, 
but is somewhat over-estimated for good solvents 
due to the small peak shift owing to the excluded volume effect. 
In summary, 
the theoretical model has revealed the transient relaxation and steady state 
features of the convective depletion dynamics of polymer solutions. 
It will be interesting to extend this model to higher-dimensional cases 
and validate the findings experimentally in the future.
%
%

\section*{Acknowledgment}
This work was supported in part by KAKENHI 
from the Ministry of Education, 
Culture, Sports, Science and Technology of Japan, 
and by the U.S. NSF under Grant No. CMMI-0952646.

\appendix

\section{Equilibrium Profile in Good Solvents}
\label{sec:Appdx:Good_Sovent}

The equilibrium profile of polymer chains in a good solvent 
is described by Eq.(\ref{eqn:zeroth_order_equation}). 
Multiplying Eq.(\ref{eqn:zeroth_order_equation})
by $d\varphi_\ro/d x$ and then integrating it once, we obtain
\begin{equation}
   {\ell^2 \over 2L^2} 
   \Bigr ( {d \varphi_\ro \over d{ x}} \Bigr )^2
 + {1 \over 2} \mu_\ro(0) \varphi_\ro^2 
 - {1 \over 4}  v         \varphi_\ro^4 = C,
\label{eqn:Appdx:mu_at_equilibrium_in_good_solvent_2}
\end{equation}
where 
$v=(1-2\chi)\bar\phi_\rp$, the constant 
$C=\varphi_{\rm m}^2\mu_\rp(0)/2 - \varphi_{\rm m}^4 {v }/4$
is determined by 
$d\varphi_\ro/dx|_{x=1/2}=0$ 
and $\varphi_\ro(1/2)=\varphi_{\rm m}$ 
at the middle point.
Therefore, Eq.(\ref{eqn:Appdx:mu_at_equilibrium_in_good_solvent_2}) 
becomes 
\begin{equation}
    \biggr ( {d  \tilde \varphi_\ro \over d { { \tx}}} \biggr )^2
= (1 - \tilde \varphi_\ro^2 )( 1 - k^2  \tilde \varphi_\ro^2 ),
\label{eqn:Appdx:varphi_equation_in_good_solvent_2}
\end{equation}
where $ \tilde \varphi_\ro \equiv \varphi_\ro/\varphi_{\rm m}$,  
$\tilde x= x \varphi_\Rm \sqrt{v L^2/2k^2\ell^2}$,
and $k^2$ is defined as 
\begin{equation}
k^2 = {v \over 2\mu_\ro(0)/\varphi_\Rm^2  - v }.
\label{eqn:Appdx:k2}
\end{equation}
Integrating Eq.(\ref{eqn:Appdx:varphi_equation_in_good_solvent_2})
yields 
\begin{equation}
 \tilde \varphi_\ro(\tx) = \sn(\tx, k), 
\label{eqn:Appdx:varphi_elliptic}
\end{equation}
where $\sn(x,k)$ is the Jacobi elliptic integral. 
From $\tilde \varphi_\ro=1$ at 
$x=1/2$ we find
\begin{equation}
\varphi_\Rm = { 2 \ell K(k) \over L } \sqrt{ 2k^2 \over v },  
\label{eqn:varphi_m}
\end{equation}
where $K(k)$ is the complete elliptic integral defined as
\begin{equation}
K(k) = \int_0^1 { dt \over \sqrt{(1-t^2)(1-k^2 t^2)} }. 
\label{eqn:ttL}
\end{equation}
From the normalization condition 
$\int_0^{1/2} \phi_\ro(x)dx =1/2$, 
we have
\begin{eqnarray}
{8k^2 \ell^2 K(k) \over  v L^2 }
\int_0^{K(k)} \sn^2 ( \tx, k )
d\tx = 1.
\label{eqn:normalization_for_phi_3}
\end{eqnarray}
Based on the properties of elliptic integrals it follows 
\begin{equation}
K(k) \bigr [ K(k)-E(k) \bigr ] = {v L^2 \over 8 \ell^2}, 
\label{eqn:eq_to_obtain_k}
\end{equation}
where $E(k)$ is the second kind complete elliptic integral,  
expressed as
\begin{equation}
E(k) = \int_0^{\pi/2} \sqrt{1 - k^2 \sin^2 \varphi~}~d\varphi.
\end{equation}
From Eqs.(\ref{eqn:Appdx:k2}) and (\ref{eqn:varphi_m}) we obtain 
\begin{equation}
\mu_\ro(0) = {4 \ell^2 K^2(k)\over L^2}(k^2 + 1) . 
\label{eqn:mu(0)}
\end{equation}
In summary, 
for a given $v$ and $L$
we can determine $k$ from Eq.(\ref{eqn:eq_to_obtain_k}), 
and then evaluate $\varphi_{\rm m}$ and $\mu_\ro(0)$ 
from (\ref{eqn:varphi_m}) and (\ref{eqn:mu(0)}), respectively.
%
%


                          \end{document}